%% file: main.tex
\documentclass[a4paper,11pt]{article}
\usepackage{jinstpub} 
\usepackage{lineno}
\usepackage{tabularx}
\usepackage[dvipsnames]{xcolor}
\usepackage{booktabs}
\usepackage{subfig}

\title{The LED calibration systems for the mDOM and D-Egg sensor modules of the IceCube Upgrade: \newline \Large Design, production, testing and use in module calibration}
\collaboration{\includegraphics[height=17mm]{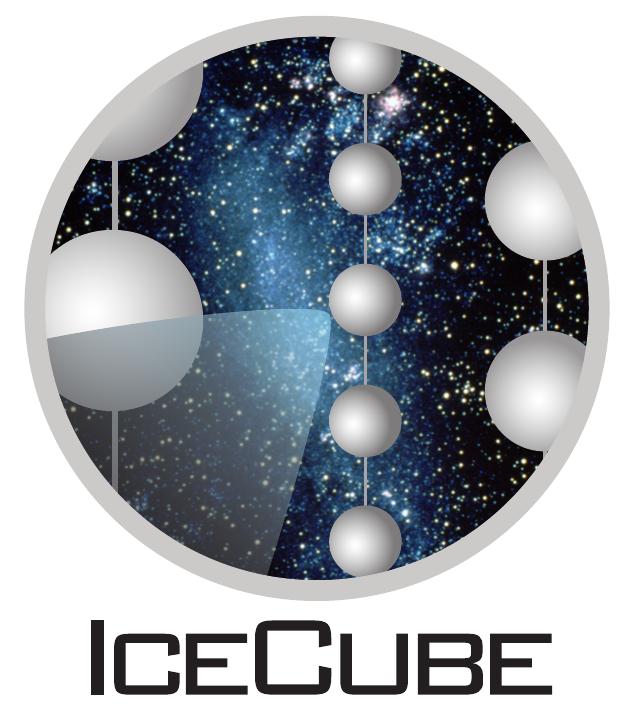}\\[6pt]IceCube collaboration}
\input{authorlist.tex}

\emailAdd{analysis@icecube.wisc.edu}

\abstract{
The IceCube Neutrino Observatory, instrumenting about 1\,km$^3$ of deep, glacial ice at the geographic South Pole, is due to be enhanced with the IceCube Upgrade. The IceCube Upgrade, to be deployed during the 2025/26 Antarctic summer season, will consist of seven new strings of photosensors, densely embedded near the bottom center of the existing array. Aside from a world-leading sensitivity to neutrino oscillations, a primary goal is the improvement of the calibration of the optical properties of the instrumented ice. These will be applied to the entire archive of IceCube data, improving the angular and energy resolution of the detected neutrino events. For this purpose, the Upgrade strings include a host of new calibration devices. Aside from dedicated calibration modules, several thousand LED flashers have been incorporated into the photosensor modules. We  describe the design, production, and testing of these LED flashers before their integration into the sensor modules as well as the use of the LED flashers during lab testing of assembled sensor modules.
}


\begin{document}
\maketitle
\flushbottom

\section{Introduction}
\label{sec:intro}

  \begin{figure}[!ht]
  \centering
    \subfloat[The red dots indicate the positions of the seven IceCube Upgrade strings within the IceCube high-energy array and its sub-array DeepCore.]{\includegraphics[width=0.53\textwidth]{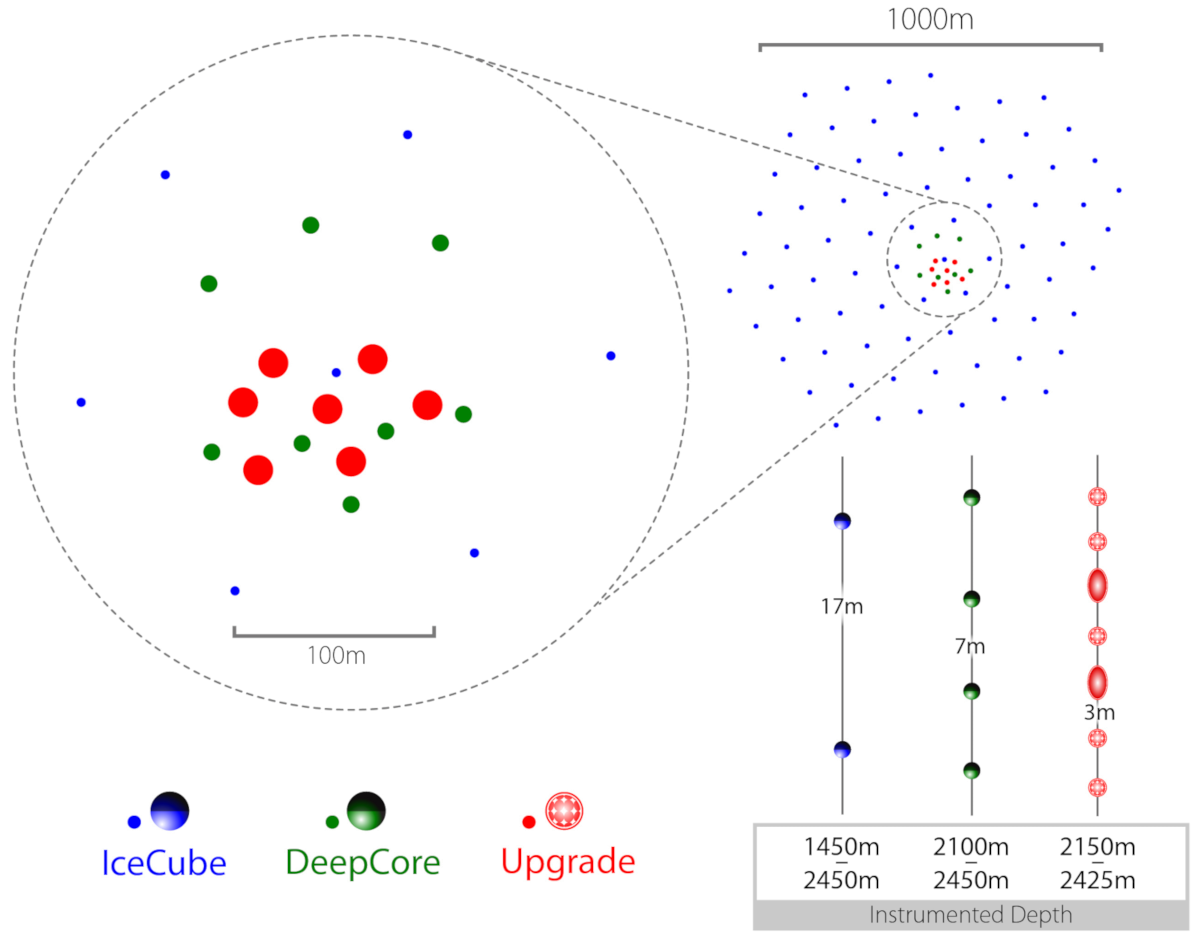}}
    \hfill
    \subfloat[The subdivision of the vertical geometry into a shallow calibration region, the depth dedicated to neutrino (oscillation) physics as well as an exploratory deep-ice region below the depth instrumented by Gen1.]{\includegraphics[width=0.45\textwidth]{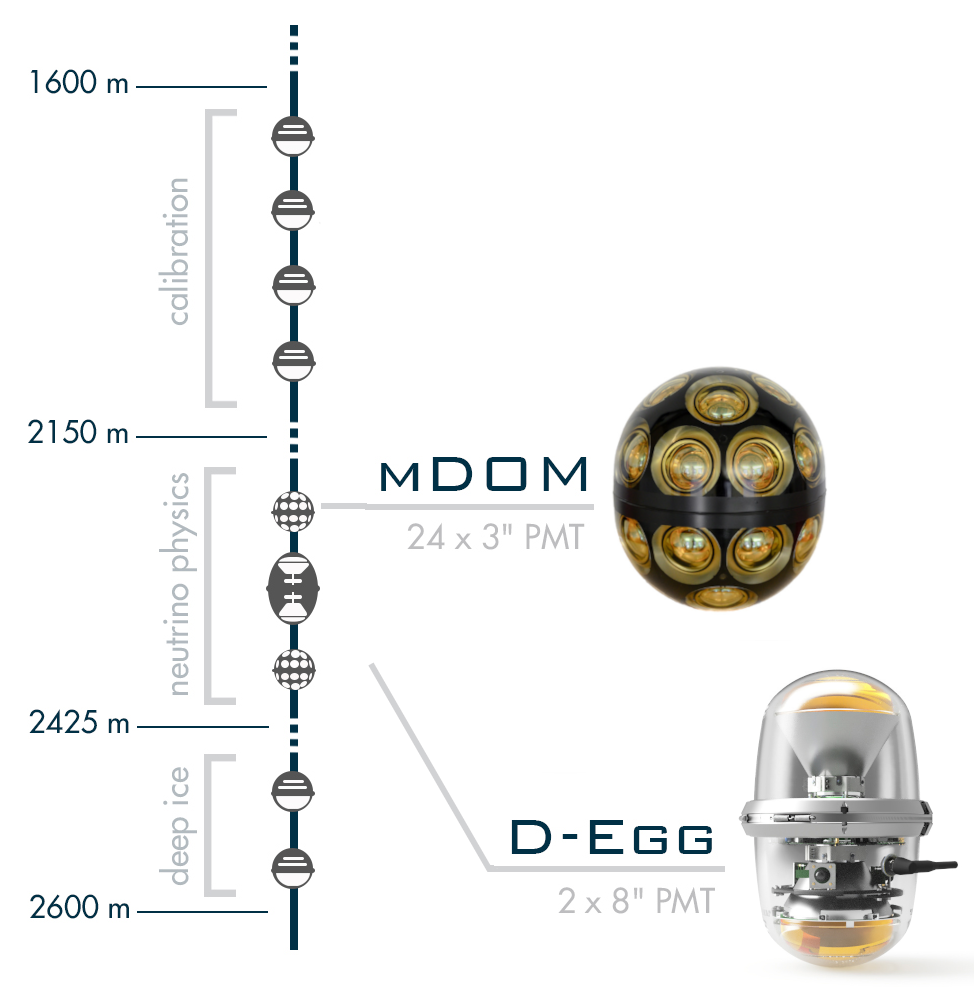}}

    \caption{The array geometry of the IceCube Upgrade.}\label{fig:UpgradeGeo}
  \end{figure}

The IceCube Neutrino Observatory \cite{detector:paper} is a Cherenkov telescope that instruments a cubic kilometer of deep, glacial ice at the geographic South Pole. The original IceCube detector, referred to as Gen1 throughout this paper, consists of 86 cables called "strings", each instrumented with 60 optical sensors called Digital Optical Modules (DOMs).
Each DOM consists of a 10-inch photomultiplier tube (PMT) and all the required readout electronics contained in a spherical glass pressure housing \cite{DOM:paper, pmt:paper}. Detector construction was completed in 2011.

As a highlight among its diverse scientific topics, IceCube has since discovered a flux of astrophysical neutrinos \cite{HESE,Abbasi2021,Abbasi2022, cascadepaper} and started to associate sources with this flux \cite{2018TXS,2022NGC,doi:10.1126/science.adc9818}. Through the study of atmospheric neutrinos with the more densely instrumented DeepCore subarray, IceCube further achieves competitive measurements of neutrino oscillation parameters \cite{PhysRevLett.120.071801, Abbasi2023}. 
As the experiment continues to accumulate lifetime, systematic uncertainties in the properties of the detector, both ice and instrumentation, begin to dominate over the statistical uncertainty for many analyses. 

Understanding the optical properties of the instrumented glacier is particularly challenging as it can only be studied in situ. For this purpose, the Gen1 DOMs are each equipped with 12 calibration LEDs. These can inject light with known intensity and timing profiles into the ice, to be detected by the surrounding array of PMTs. Using these data, the modeling of ice optical properties has recently been updated to include the effect of light deflection due to the birefringent microstructure of the ice \cite{tc-18-75-2024} as well as detailed maps of isochron undulations \cite{TiltICRC}. There remain uncertainties about the optics of the ice, such as the properties of refrozen water in the drill holes \cite{Rongen2016, Chirkin2021} and the distribution of scattering angles. 

To further improve the sensitivity to neutrino oscillation physics and address remaining calibration uncertainties, the detector will be supplemented by seven additional instrumentation strings in the austral summer of 2025/26. The layout of this so-called IceCube Upgrade \cite{IceCubeUpgrade} can be seen in figure \ref{fig:UpgradeGeo}. The new strings will be deployed within the DeepCore footprint, thus lowering the energy threshold through an increase in photocathode density. Along each string, the vertical distance between modules is further reduced from 17\,m in IceCube and 7\,m in DeepCore to 3\,m in the Upgrade for most depths. In contrast to IceCube, the Upgrade does not feature a single sensor design but has two primary module types interleaved on each string. The D-Egg \cite{Abbasi2023Degg} (see figure \ref{fig:DEggModule}) features two 8" PMTs in an elongated pressure vessel. The mDOM \cite{Classen2021} (see figure \ref{fig:mDOMModule}) features 24 3" PMTs distributed nearly isotropically.
Additional engineering modules (such as the LOMs \cite{Shimizu2021} and WOMs \cite{BastianQuerner2022}), as well as several device types dedicated to specific calibration aspects (such as the POCAM \cite{Henningsen2020}, the acoustics system \cite{Gnther2023} and the Pencil Beam \cite{Rongen2021WhatTo}), are strategically distributed throughout the array.

\pagebreak

Like the IceCube Gen1 DOM, the Upgrade mDOMs and D-Eggs feature calibration LEDs that will provide data for future improvements to the ice optical modeling. To ensure a homogeneous dataset, the LED systems for both module types have been developed in tandem and have been tested to the same specifications with equivalent setups. Here, we describe the design, production, and testing of this LED calibration system.

\section{Design of the LED calibration system}

\subsection{Design requirements}

The IceCube Gen1 DOMs use 405\,nm LEDs in order to approximate the typical wavelength of detected Cherenkov photons from particle interactions in ice \cite{Aartsen2013}. The LEDs in the Gen1 DOMs are driven by high-speed MOSFET drivers with variable width and amplitude settings. This allows for intensities between $10^6$ and $1.4\cdot10^{11}$ photons per pulse. In the Gen1 design, the highest intensity is only obtainable at the longest pulse width, which equates to a rectangular pulse of 70\,ns FWHM. The lower intensities allow for shorter pulses, with the fastest configuration resulting in a nearly Gaussian pulse of 6\,ns FWHM. 

For the LED flashers in the mDOM and D-Egg modules of the IceCube Upgrade, the same wavelength has been chosen, with light sources in special devices, such as the POCAM, providing data at additional wavelengths. 
Due to the reduced distance between emitters and receivers in the Upgrade compared to the Gen1 array, the intensity requirement has been relaxed for the Upgrade. In order to resolve small changes in the received arrival time distributions resulting from effects such as changes to the assumed scattering function, the Upgrade requires shorter LED pulse durations. Following simulation studies, the desired intensity range has been set from $5\cdot10^6$ to $10^9$ photons per pulse, while the pulse width should not exceed 10\,ns FWHM, even at maximum brightness.\\

\renewcommand{\arraystretch}{1.2}
\begin{table}
    \centering
    \begin{tabularx}{\textwidth}{p{0.29\textwidth}|p{0.67\textwidth}}
        \toprule
        Requirement & Description  \\
        \midrule
        Operational temperature  & Meet specifications between -30$^{\circ}$C\newline (accounting for module self-heating) and +27$^{\circ}$C. \\
        Brightness range & $5\cdot 10^6$ to $10^9$ photons per pulse. \\
        Emission spectrum & Central value of $405\pm10$\,nm with a FWHM below 30\,nm.\\
        Angular distribution &  Approximately Gaussian in each degree of freedom with standard deviations not exceeding 15 degrees. \\
        Time profile & FWHM $\leq$ 7\,ns for dim settings and $\leq$10\,ns at maximum brightness.\\
        Accuracy of emission axis & Each LED axis shall be aligned within a 5-degree solid angle tolerance of its nominal design direction.   \\
        Controllability of brightness & The LED intensity shall be configurable to within $\pm 50\%$ of any target value in its dynamic range. \\  
        Consistency of brightness & The per pulse output of each LED shall be consistent to within 10\% standard deviation for a duration of 30 minutes. \\     
        \bottomrule
    \end{tabularx}
    \caption{Selected design requirements for the LED calibration system.}
    \label{tab:requirements}
\end{table}

In 2013 it was discovered that light propagation in the glacial ice features strong anisotropies \cite{ICRC_anisotropy}. Satisfactory calibration of this effect \cite{tc-18-75-2024} requires flashing each LED in the detector individually while knowing the emission axis of each LED. In the IceCube Gen1 DOMs the LED orientations were provided by simply bending the LED leads through the use of a jig, resulting in an unknown orientation precision. For the LEDs in the Upgrade modules, we require well-defined mechanical solutions that guarantee that each LED axis points to within a 5-degree solid angle of its nominal design direction.

Due to fluctuations in LED and pulse driver performance, as well as temperature dependencies, the light output for a given intensity configuration will differ between LEDs. The temperature of the instrumented ice in particular features a depth gradient, where the temperature increases from $-40^{\circ}$C at 1600\,m depth to $-20^{\circ}$C at 2400\,m depth \cite{Price2002}. To enable predictable in situ operation for all devices, resulting in sufficient photon statistics while avoiding saturation of the receivers, we require that the intensity of each LED can be configured to within $\pm 50\%$ of any target value across
the full dynamic range based only on prior lab calibration.

A summary, including further requirements on the operational temperature range, the angular emission profile, and the long-term consistency of brightness, can be found in table \ref{tab:requirements}.

\subsection{Common schematic and parts selection}

  \begin{figure}[!ht]
  \centering
    \subfloat[Simplified schematic of a single LED element.\label{subfig-1:schematic}]{%
      \includegraphics[width=0.85\textwidth]{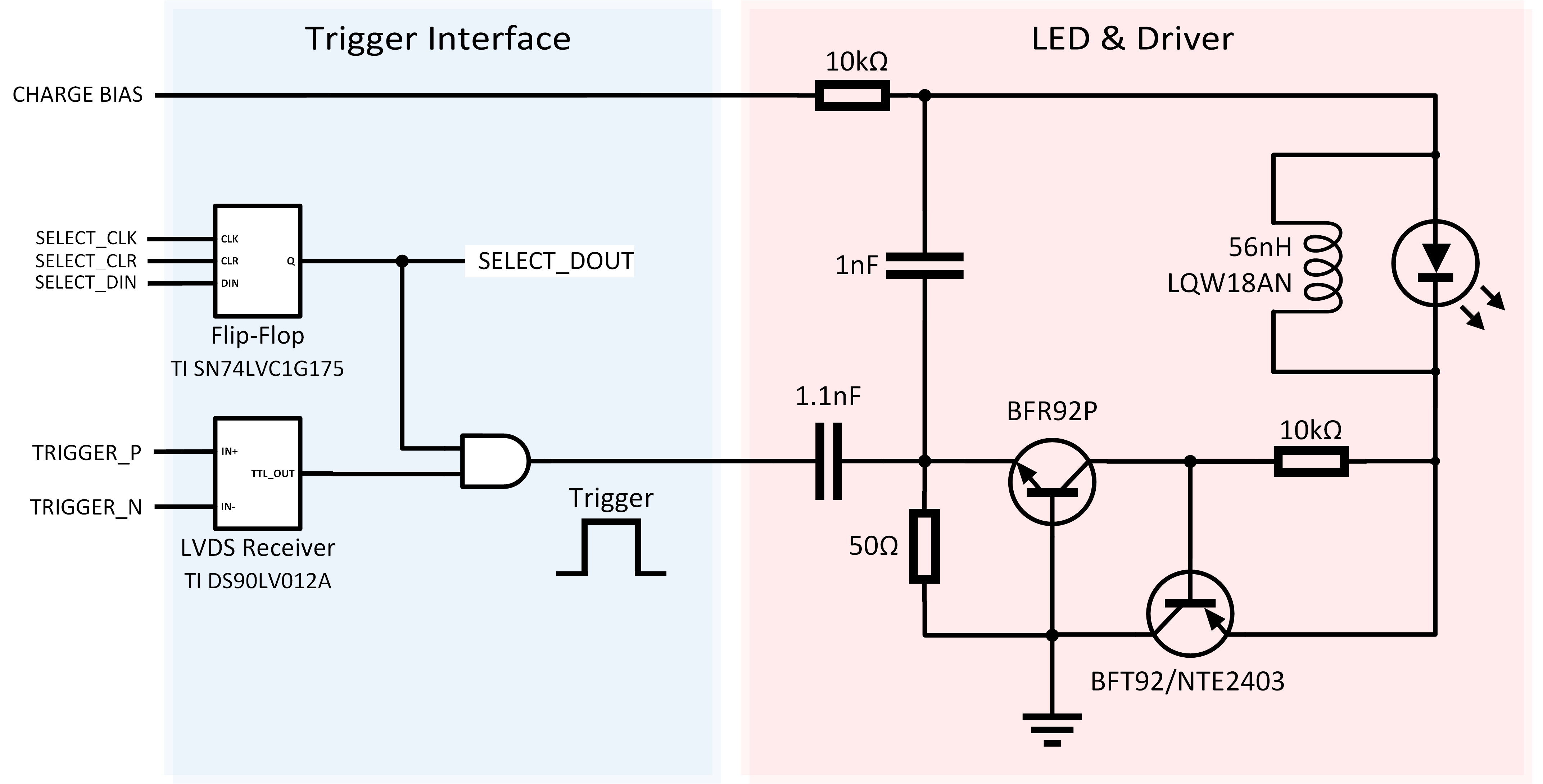}
    }\\
    \subfloat[Block diagram of multiple LED elements forming a flasher daisy chain.\label{subfig-2:blockdiagram}]{%
      \includegraphics[width=0.85\textwidth]{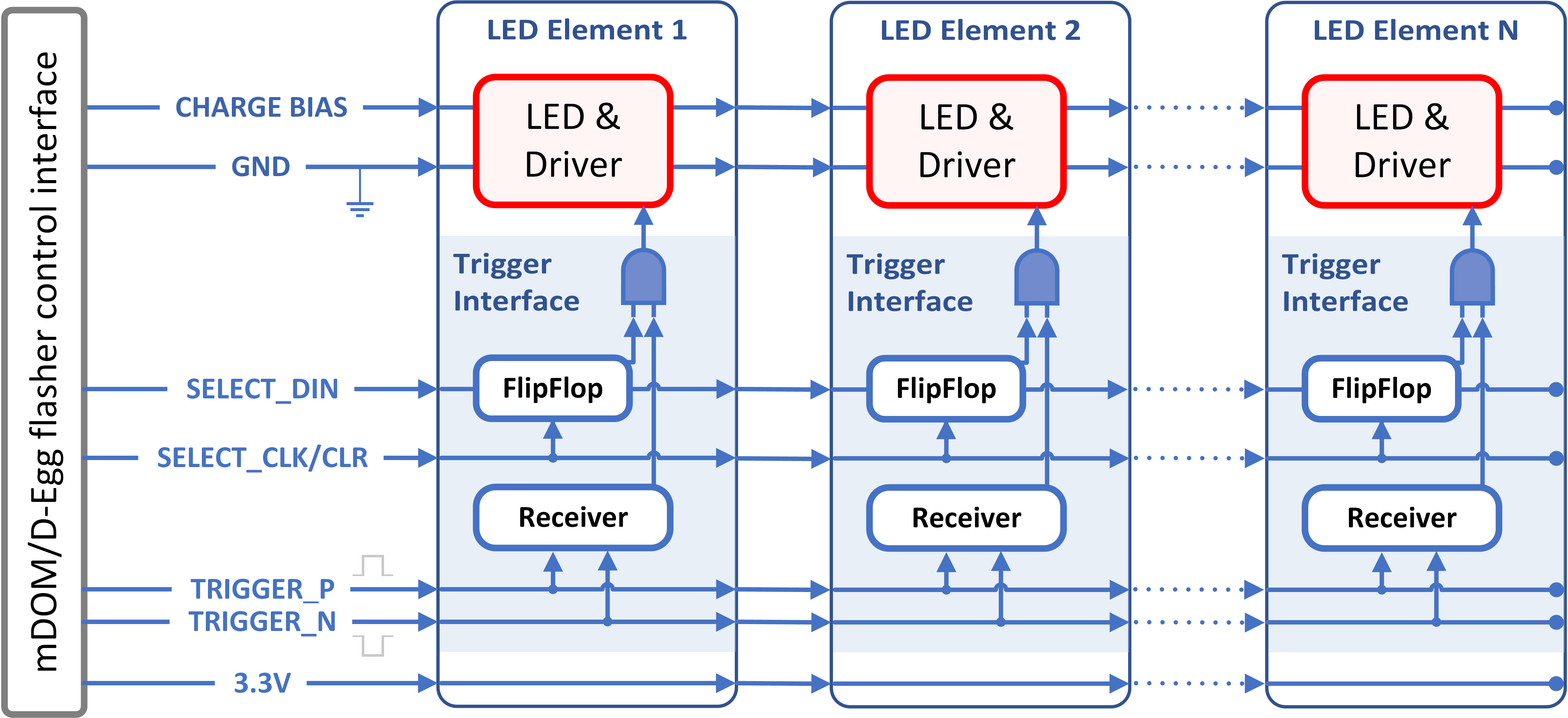}
    }
    \caption{Schematic and block diagram of the LED flasher systems. The mDOM and D-Egg systems both conform to this design but differ in their implementation due to different mechanical requirements.
    Each mDOM features two daisy chains with 5 LED elements, where each LED element is realized as a separate PCB. In the D-Eggs the 12 LED elements are arranged in one daisy chain on one large PCB.}\label{fig:design}
  \end{figure}

The pulse driver for each LED is based on a design by Kapustinsky in 1985 \cite{Kapustinsky1985}. It has been chosen for its robustness and ease of miniaturization and is also used by other devices in the field, such as the POCAM \cite{Henningsen2020} and the KM3NeT nanobeacons \cite{Aiello2022}. The design is based on the triggered discharge of a small capacitor through a pair of RF transistors that form a thyristor-like element (see the schematic in figure \ref{subfig-1:schematic}). An optional inductor parallel to the LED cuts off the trailing edge of the light pulse. The intensity can be adjusted by varying the bias voltage into the discharge capacitor. 

To meet the requirements listed in table \ref{tab:requirements}, several LED types in combination with different discharge capacitances, inductances, and maximum bias voltages were tested. The chosen design employs a Roithner XRL-400-5O 5\,mm LED\footnote{\url{http://www.roithner-laser.com/datasheets/led_div/xrl-400-5o.pdf}}, a 1\,nF discharge capacitor, a 56\,nH inductance and a maximum bias voltage of 15\,V. All control and trigger signals are provided by the module's mainboard. The bias voltage is regulated via a 16-bit digital-to-analog converter, offering ample resolution for fine-tuning.

To minimize the number of interfaces required to operate multiple LEDs per module, several LED elements are bundled into a daisy chain as seen in figure \ref{subfig-2:blockdiagram}. Each LED module comprises the LED, a pulse driver, and a trigger interface. Individual LEDs along the daisy chain can be enabled or disabled through a trigger word, which is distributed along the chain via flip-flops. A common bias voltage and differential trigger signal are distributed to all LED elements. Since the cable delay increases down the daisy chain, this results in the emission time being offset by roughly one nanosecond between each LED. Also note that the LED driver triggers on the falling edge of the positive trigger pulse from the LVDS receiver.

\subsection{Implementation in the mDOM}

  \begin{figure}[!ht]
  \centering
    \subfloat[Each flasher comprises a 25\,mm by 15\,mm PCB and a single outward pointing LED. Five of these LED elements wired in series via ribbon cables form a flasher daisy chain.\label{subfig-1:mDOMLEDelement}]{%
      \includegraphics[width=0.4\textwidth]{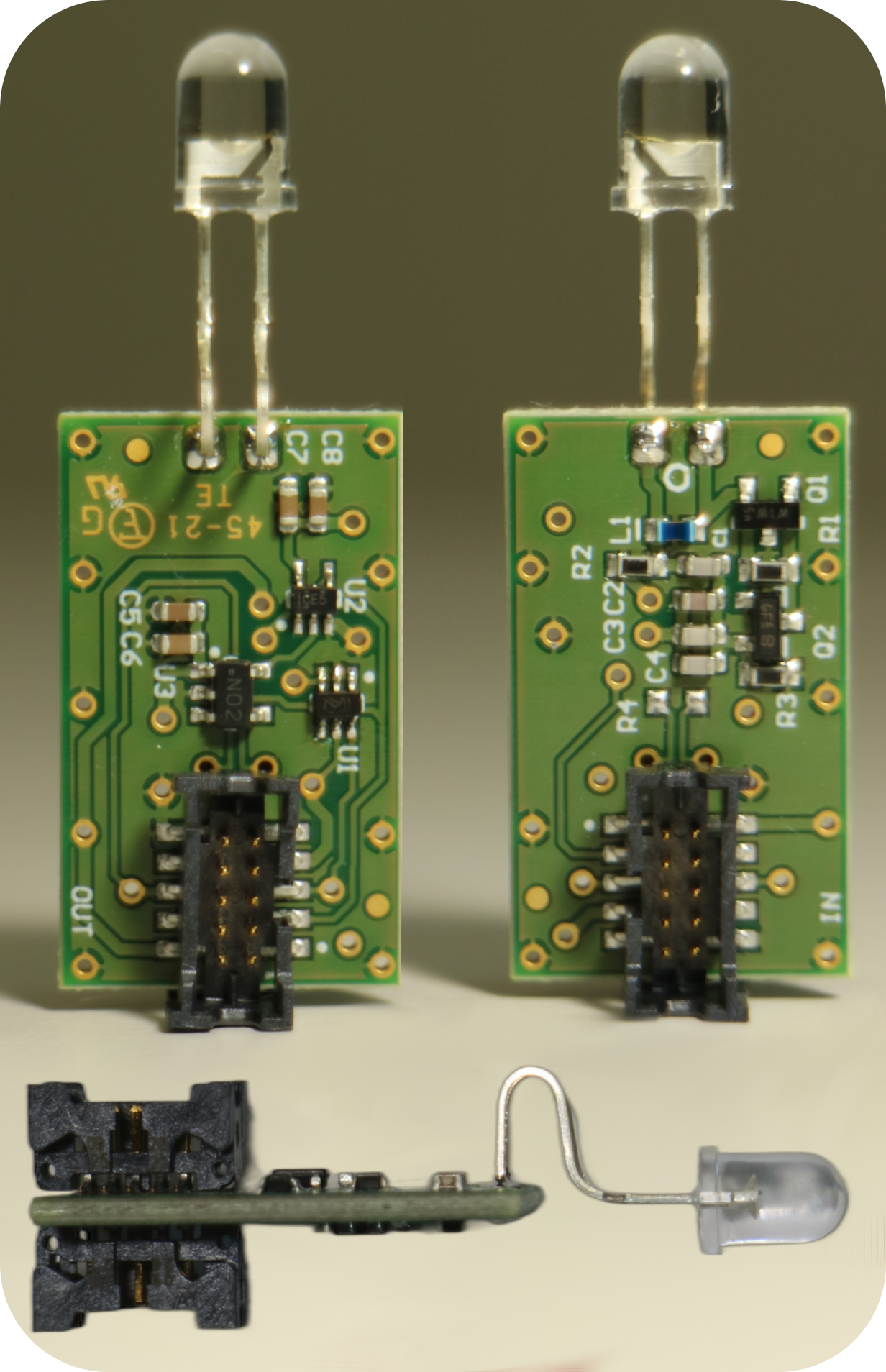}
    }\hfill
    \subfloat[Photo of an assembled mDOM. One flasher daisy chain is integrated into the black support structure of each hemisphere. Eight LEDs point outward at elevation angles of $\pm 29^{\circ}$, and two LEDs point vertically upward/downward. The LED is separated from the optical gel, which glues the support structure into the glass pressure vessel, via a small glass window. \label{fig:mDOMModule}]{%
      \includegraphics[width=0.55\textwidth]{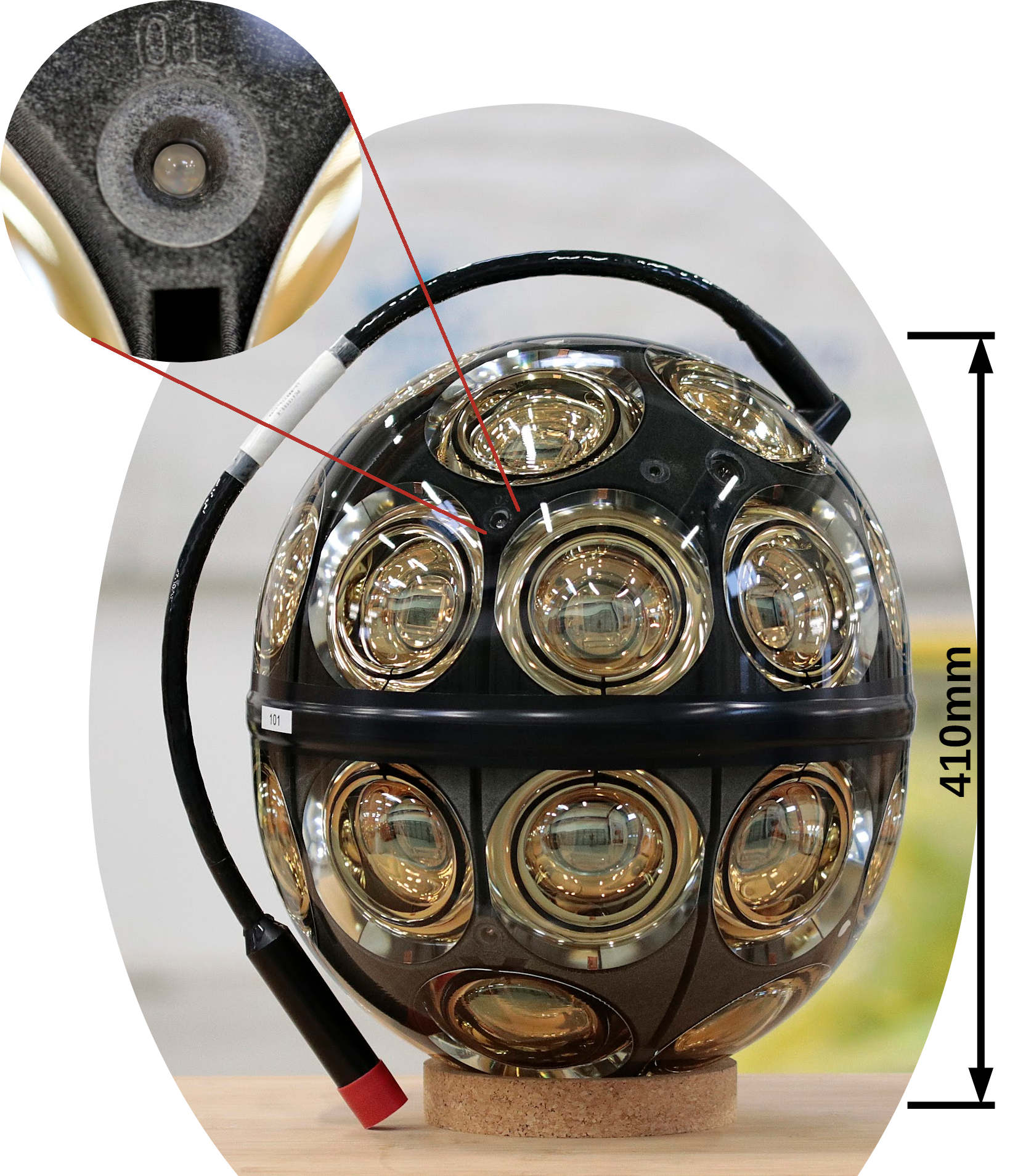}
    }
    \caption{Implementation of the LED calibration system within the mDOM.}
  \end{figure}

Due to the large number of PMTs, space in the mDOMs is limited and highly segmented. Thus the LED elements have been realized as individual printed circuit boards (PCBs), as shown in figure \ref{subfig-1:mDOMLEDelement}. Five of these are arranged in a daisy chain connected with flexible ribbon cables. One daisy chain is installed in each mDOM hemisphere and connects to the calibration extension board (called mDAB) of the mDOM mainboard. Four LEDs in each hemisphere point at elevation angles of $\pm29^{\circ}$ and one LED points nearly vertically up or down (81$^{\circ}$ elevation). The LEDs of both hemispheres are aligned, resulting in four pairs of LEDs spaced 90$^{\circ}$ apart in azimuth.
The flasher PCBs are mounted within slots in the 3D-printed black support structure that also fixes the PMTs. Borosilicate glass windows of 1.75\,mm thickness are glued into recesses in the holding structure in front of the LEDs. These are necessary as the space between the support structure and the pressure vessel is filled with optical gel. Nominally the distance between the LED and the glass window is 1\,mm. To avoid uncontrolled flexure of the LED in the unlikely case of contact, the LED leads are bent into the spring-like arrangement seen in figure \ref{subfig-1:mDOMLEDelement}. The long lead length also avoids thermal damage to the LEDs during the soldering process.

\subsection{Implementation in the D-Egg}

  \begin{figure}[!ht]
  \centering
    \subfloat[The flasher daisy chain for each D-Egg is realized on one large, 115\,mm outer diameter, annular PCB. Eight LEDs point horizontally outwards. Four LEDs point vertically down. The orientation of each LED is ensured via a commercial, cup-type LED housing. \label{subfig-1:DEggPCB}]{%
      \includegraphics[width=0.58\textwidth]{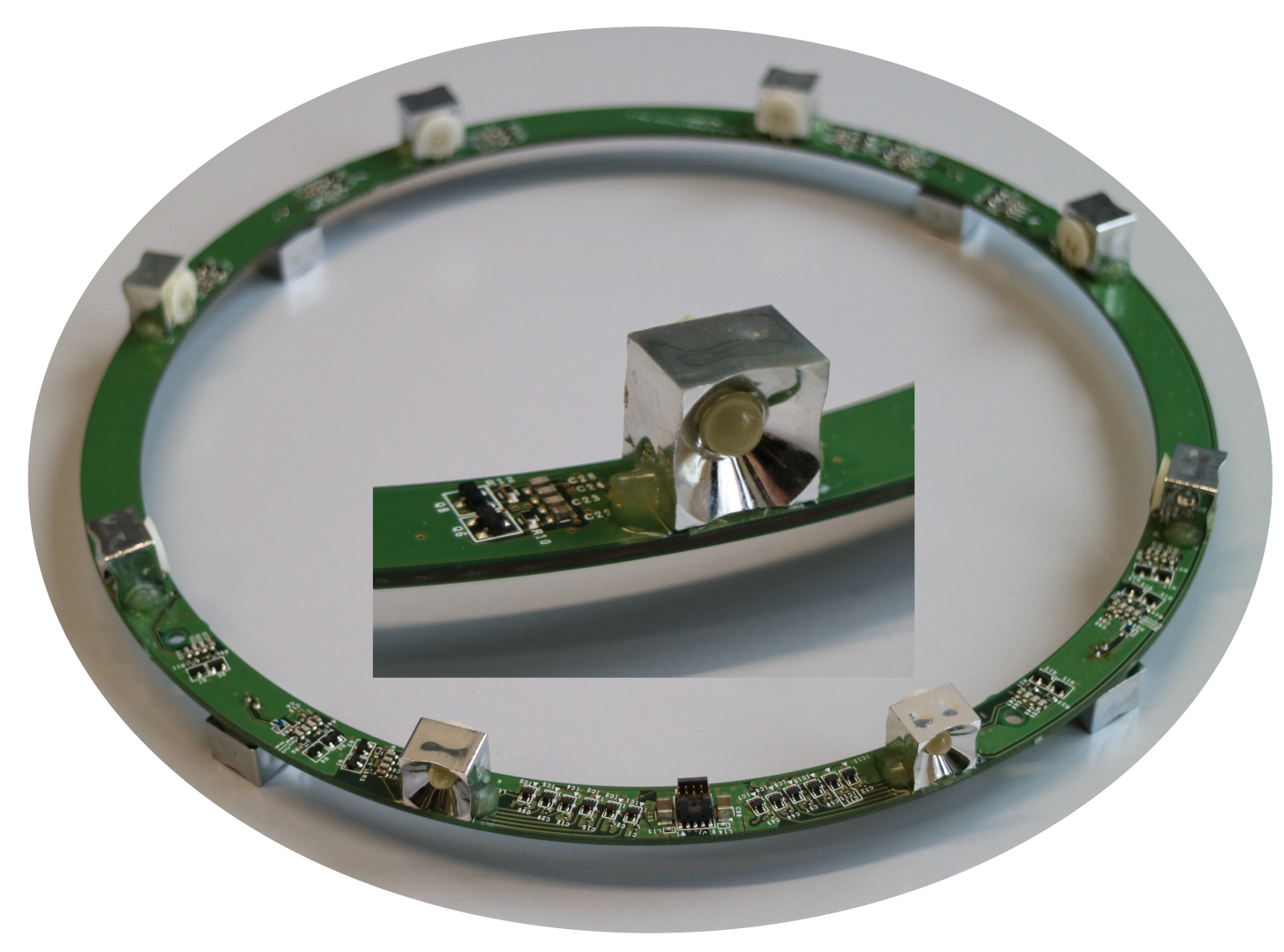}
    }\hfill
    \subfloat[CAD rendering of an assembled D-Egg. The annular PCB comprising the LED calibration system can be seen resting in the optical gel gluing the lower PMT to the glass pressure vessel. \label{fig:DEggModule}]{%
      \includegraphics[width=0.38\textwidth]{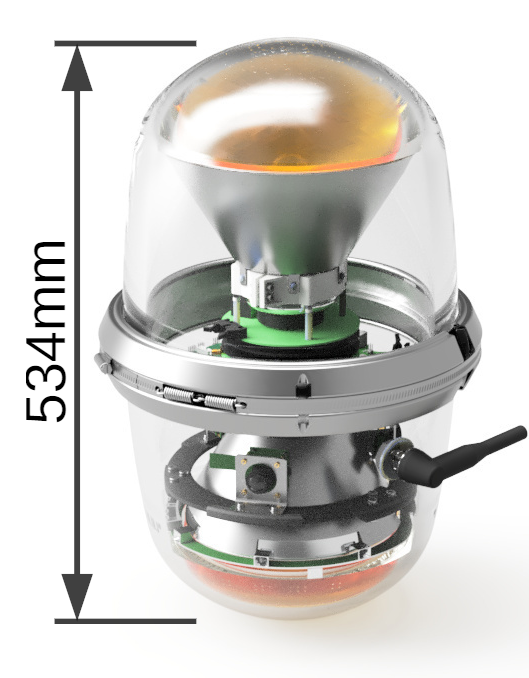}
    }
    \caption{Implementation of the LED calibration system within the D-Egg.}
  \end{figure}

The elongated pressure vessel of the D-Eggs (see figure \ref{fig:DEggModule}) offers ample space. Thus one annular PCB as seen in figure \ref{subfig-1:DEggPCB} has been designed to fit 12 LEDs. Eight LEDs point horizontally outward at 45$^{\circ}$ azimuthal increments. Four LEDs point straight down. While the downward-facing LEDs all point in the same direction, their positional offsets can be used to study the properties of the refrozen drill columns \cite{mDOMholeice}. The LEDs are placed using commercial LED mounts. The 12 LEDs form a single daisy chain and are connected to the D-Egg mainboard with a ribbon cable. During module integration, the annular flasher PCB slides past the bottom PMT and is oriented on top of the optical gel. Plastic spacers are then glued to the pressure vessel, to fix the PCB in place.

\section{Design verification}

\begin{figure}[ht]
  \centering
    \subfloat[Temporal light curves measured for one mDOM LED element at two different bias voltages, measured as a histogram of individual photon delays to a trigger signal using an avalanche photodiode.\label{subfig-1:timing}]{%
      \includegraphics[width=0.48\textwidth]{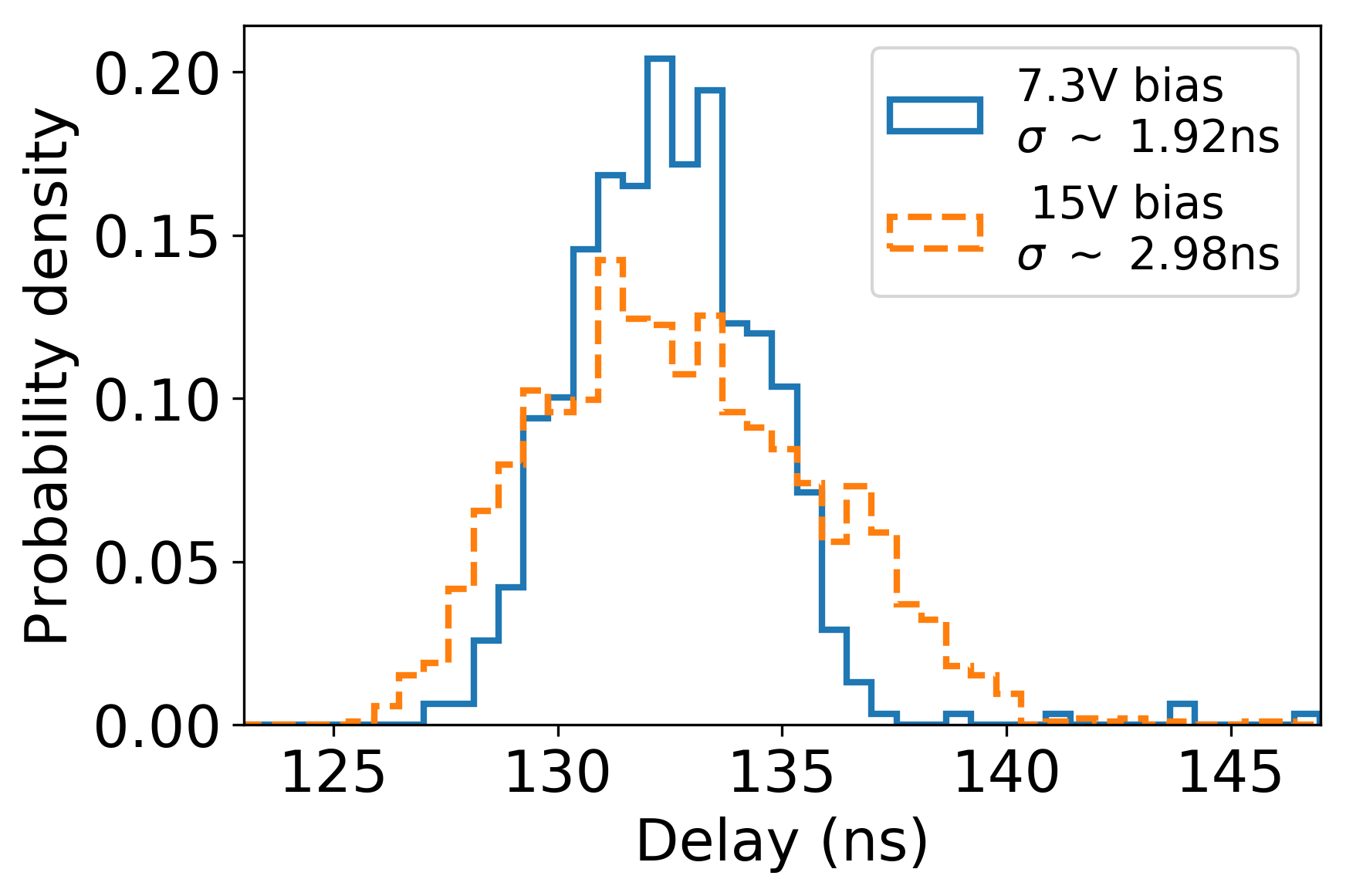}
    }\hfill
    \subfloat[Angular emission profile measured for the selected LED operated at continuous, bright emission and measured with a photodiode at a distance of one meter. \label{subfig-2:angular}]{%
      \includegraphics[width=0.48\textwidth]{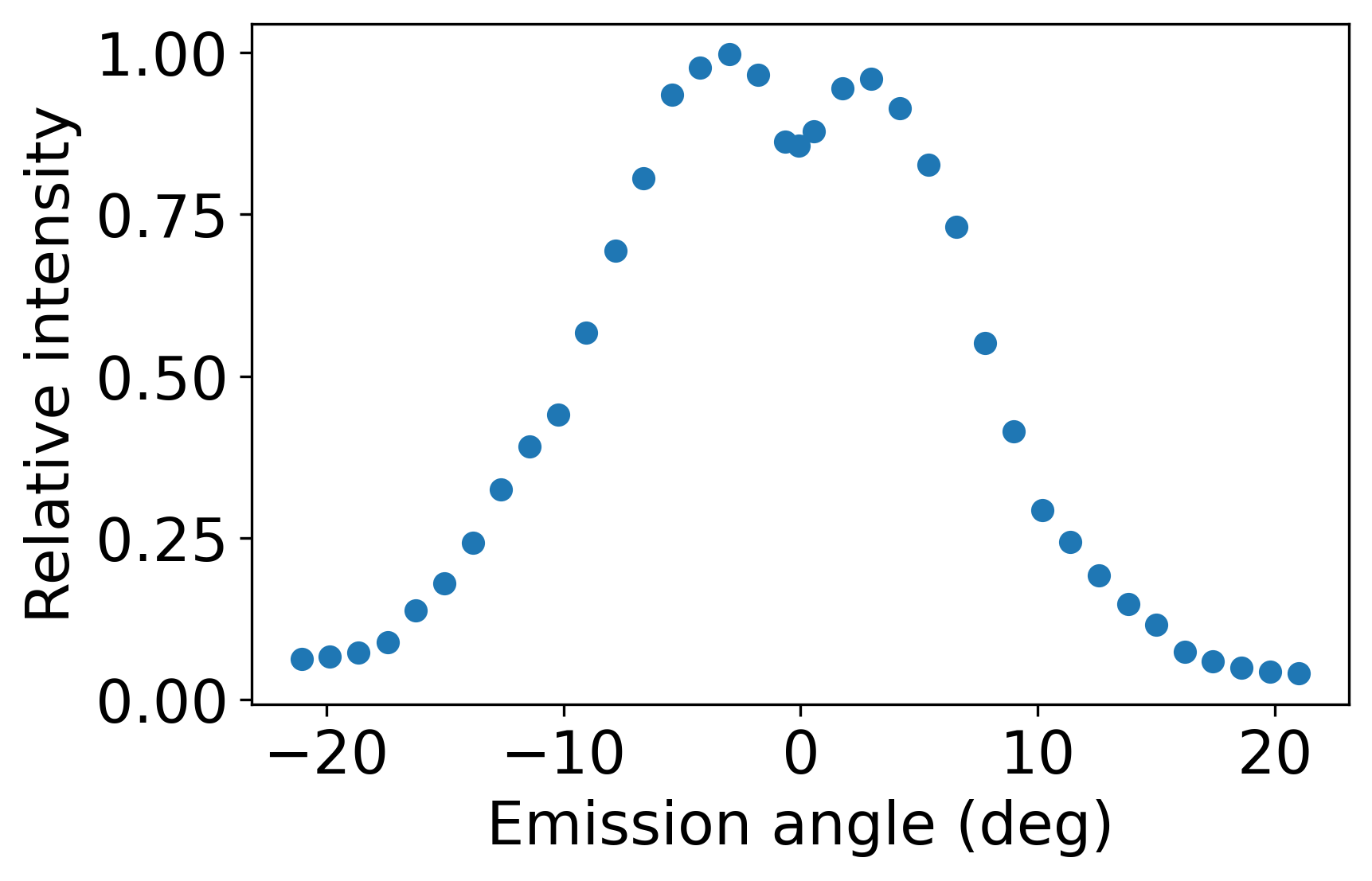}
    }
    \caption{Example plots for selected quantities evaluated during design verification.}
    \label{fig:designverification}
  \end{figure}

Before production, all design requirements (see table \ref{tab:requirements}) were verified for a prototype version of the mDOM flasher elements. Figure \ref{fig:designverification} provides two examples of measurements that have been performed in this context. 
Figure \ref{subfig-1:timing} shows the time profile of the emitted light pulses at the two ends of the intended bias voltage range. These have been measured as the distribution of timing delays between the trigger pulse going to the LED element and the arrival time of single photons at an IDQ ID100 avalanche photodiode\footnote{\url{https://www.idquantique.com/quantum-sensing/products/id100/}}, as described in \cite{Rongen2018}. To sample the arrival time distribution in an unbiased fashion, the beam is attenuated using absorptive neutral density filters to result in a detection at the avalanche photodiode in roughly 10\% of all triggers. The stated spread includes the duration of the light pulse as well as the trigger jitter and the resolution of the avalanche photodiode setup. The latter two are subdominant at approximately 100\,ps and 70\,ps, respectively. Figure \ref{subfig-2:angular} shows the angular emission profile, measured by rotating a photodiode around the LED at a distance of one meter. The small dip in intensity in the very forward region is related to the silicone resin lens encapsulating the LED die (see figure \ref{subfig-2:angular}).\\

Most performance characteristics exhibit a temperature dependence and will thus have a depth dependence within the deployed array, where the temperature increases from $-40^{\circ}$C at 1600\,m depth to $-20^{\circ}$C at 2400\,m depth \cite{Price2002}. These are mostly inconsequential (for example a 3\,nm shift in the median emission wavelength from $5^{\circ}$C to $-40^{\circ}$C), except for a roughly 40\% reduction in intensity observed at maximum bias voltages when going from room temperature to $-40^{\circ}$C. 

Additional tests have been performed for the individual implementations in the D-Egg and mDOM settings. For the mDOM mounting scheme, the pointing accuracy of the LED axis was ensured by camera inspection of 21 LED elements inserted into a cut-away version of the holding structure. Along the axis that the LED leads were bent (refer back to figure  \ref{subfig-1:mDOMLEDelement}), an average off-axis angle of $2^{\circ}$ has been measured. In the orthogonal axis, along the plane of the LED leads, the average off-axis angle was $<1^{\circ}$. No outliers beyond $5^{\circ}$ were observed.

\section{Production}

\subsection{mDOM system}

\begin{figure}[!ht]
  \centering
    \subfloat[Six panels of 220 flasher PCBs each awaiting de-panelization.\label{subfig-1:PCBpanel}]{%
      \includegraphics[height=0.32\textwidth]{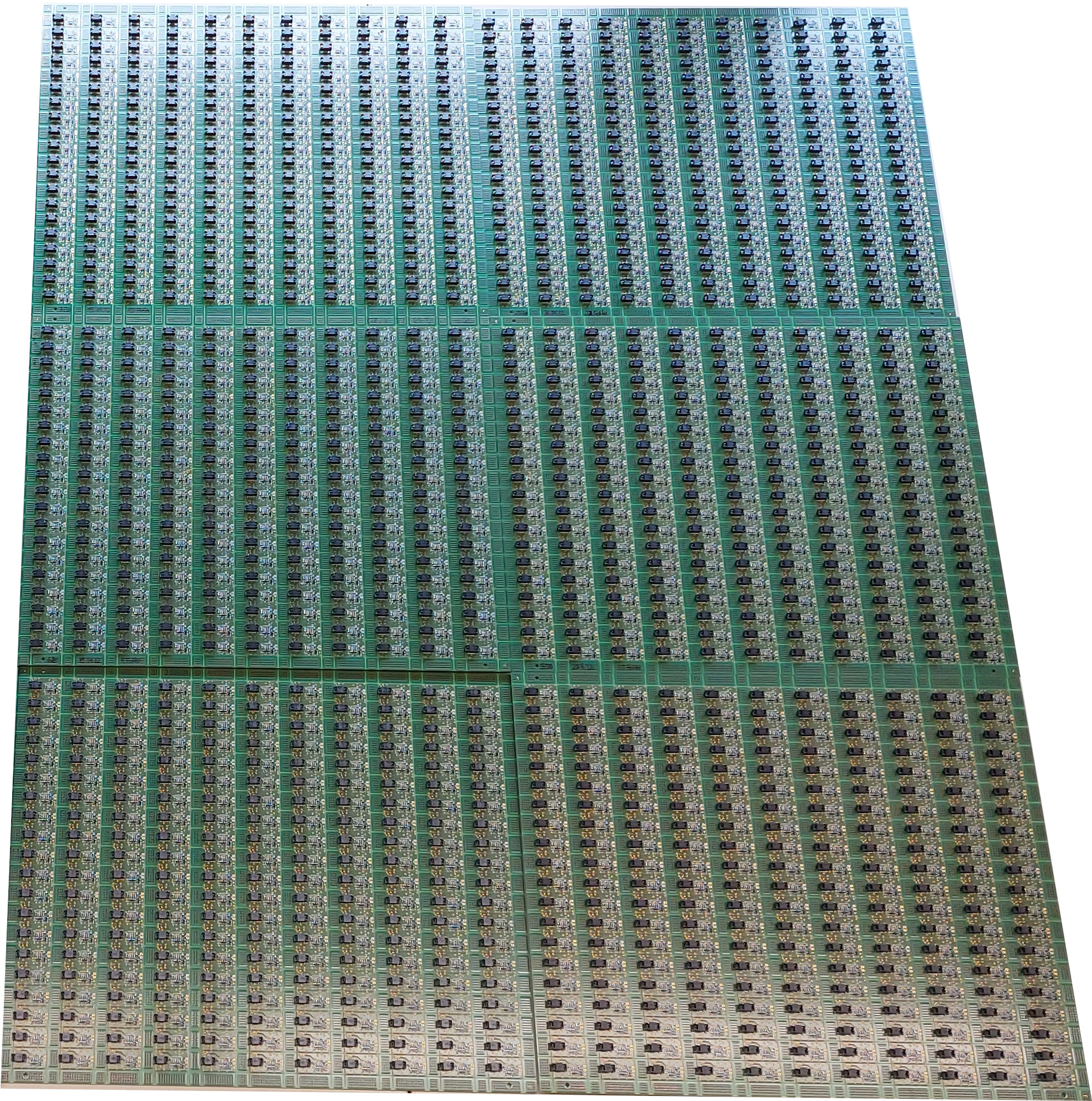}
    }\hfill
    \subfloat[Flasher in the jig used to bend the LED and solder it to the PCB.\label{subfig-2:bendingjig}]{%
      \includegraphics[height=0.32\textwidth]{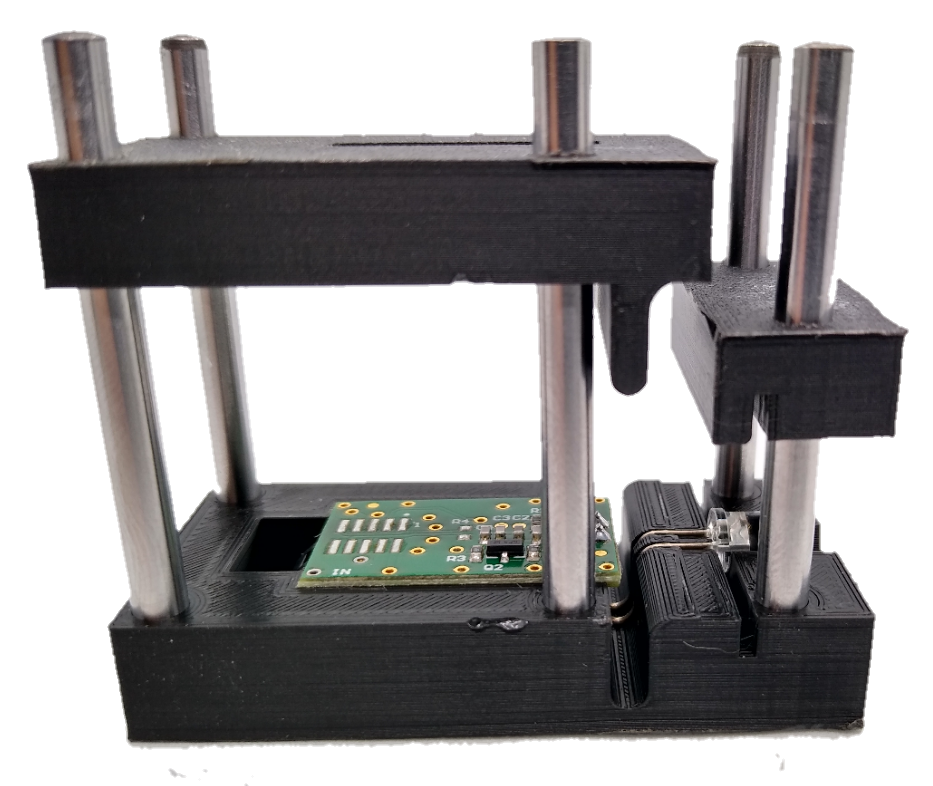}
    }\hfill
    \subfloat[One of five flashers inside its protective cover.\label{subfig-3:mDOMelement}]{%
      \includegraphics[height=0.32\textwidth]{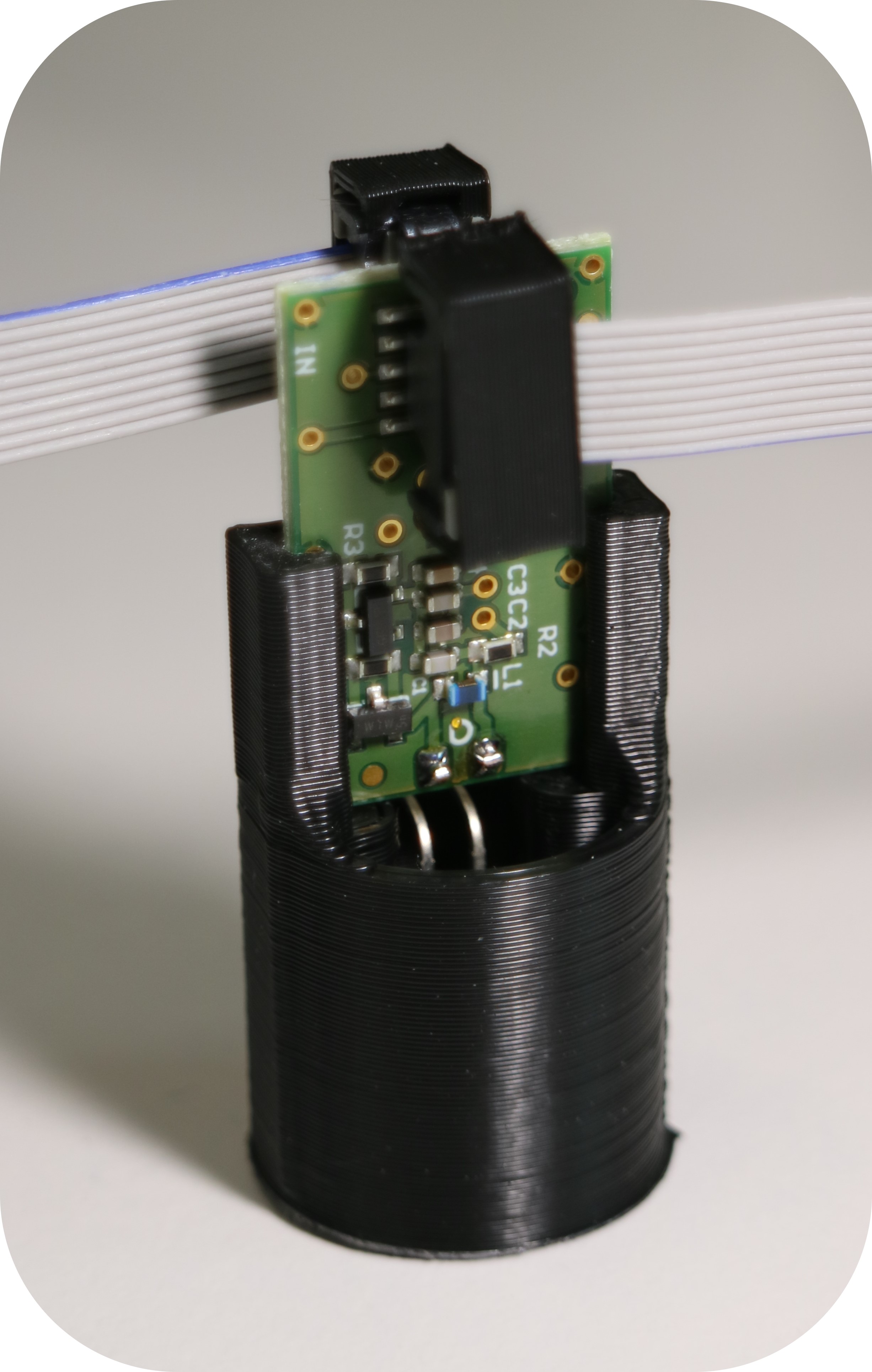}
    }
    \caption{Production cycle of an mDOM flasher daisy chain (FDC).}
  \end{figure}

The pre-populated PCBs of the mDOM flashers arrived on panels of 220 PCBs each (see figure \ref{subfig-1:PCBpanel}). To assemble one LED element, first, the LED leads were bent into the aforementioned shape (see figure \ref{subfig-1:mDOMLEDelement}). To achieve the required precision a custom jig was used, as seen in Figure \ref{subfig-2:bendingjig}. A strain relief block secures the acrylic lens cap of the LED, while a stamp presses the LED leads into a mold. With the LED still in the jig, the PCB is placed on top and the LED is manually soldered. To avoid potential bending of the LED leads during further handling and shipping, the LED element was then slotted into a 3D-printed protective cover as seen in figure \ref{subfig-3:mDOMelement}. Five prepared LED elements were connectorized to a flasher daisy chain (FDC) using ribbon cables. All connectors are further secured with custom retention clips.
Each FDC features a unique identifier, printed in clear text and as a bar code on a label on the ribbon cable going to the mDOM mainboard. 

Assembly of one daisy chain following the steps outlined above took an experienced worker roughly 20 minutes. This is in pace with the duration of acceptance testing described in section \ref{sec:testing}. 
Production and testing of $\sim$900 mDOM FDCs for 430 mDOMs, equating to over 4500 flasher elements, was completed over a time period of roughly two years.

\subsection{D-Egg system}

Due to its monolithic design, production of the D-Egg flasher system was simpler compared to the mDOM system. The annular PCBs were fully produced by an industrial contractor including the LEDs inside their reflector mounts and were ready to be tested when received.

\section{Acceptance testing}
\label{sec:testing}

For quality assurance, every LED was tested prior to integration into the modules. While the sheer number of devices to be tested prohibits an evaluation of all requirements, the basic functionality is ensured. This is particularly important due to the daisy chain design, as a malfunctioning trigger interface or broken connector on one LED element may incapacitate all LEDs further down the chain.
In addition to functionality testing, acceptance testing offers the possibility to record individual calibration constants needed for in situ operation. Acceptance testing was performed at room temperature and recorded the average per-pulse photon count at the maximum bias voltage and the bias voltage required to obtain $5\cdot10^6$ photons per pulse.

\subsection{Testing setups}
\label{sec:testsetup}

\begin{figure}[!ht]
  \centering
    \subfloat[One S2281 photodiode attached to the custom trans-impedance amplifier. Below, a custom board that digitizes the voltages coming from the amplifiers is shown.\label{subfig-1:pdsystem}]{%
      \includegraphics[width=0.32\textwidth]{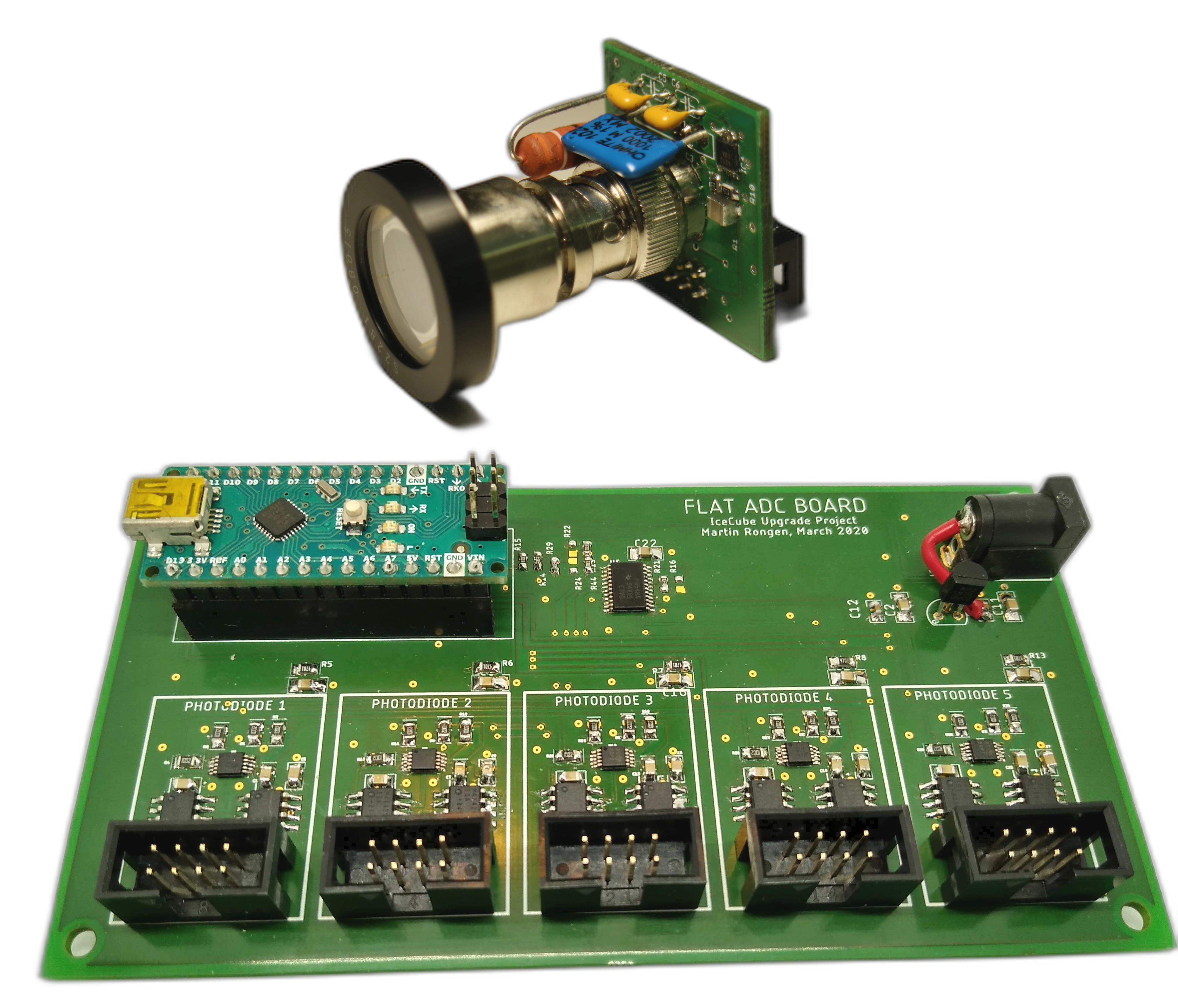}
    }\hfill
    \subfloat[Dark box containing the final acceptance testing setup for mDOM LED daisy chains. Each LED to be tested is placed in a small photodiode module. \label{subfig-2:mDOMtesting}]{%
      \includegraphics[width=0.32\textwidth]{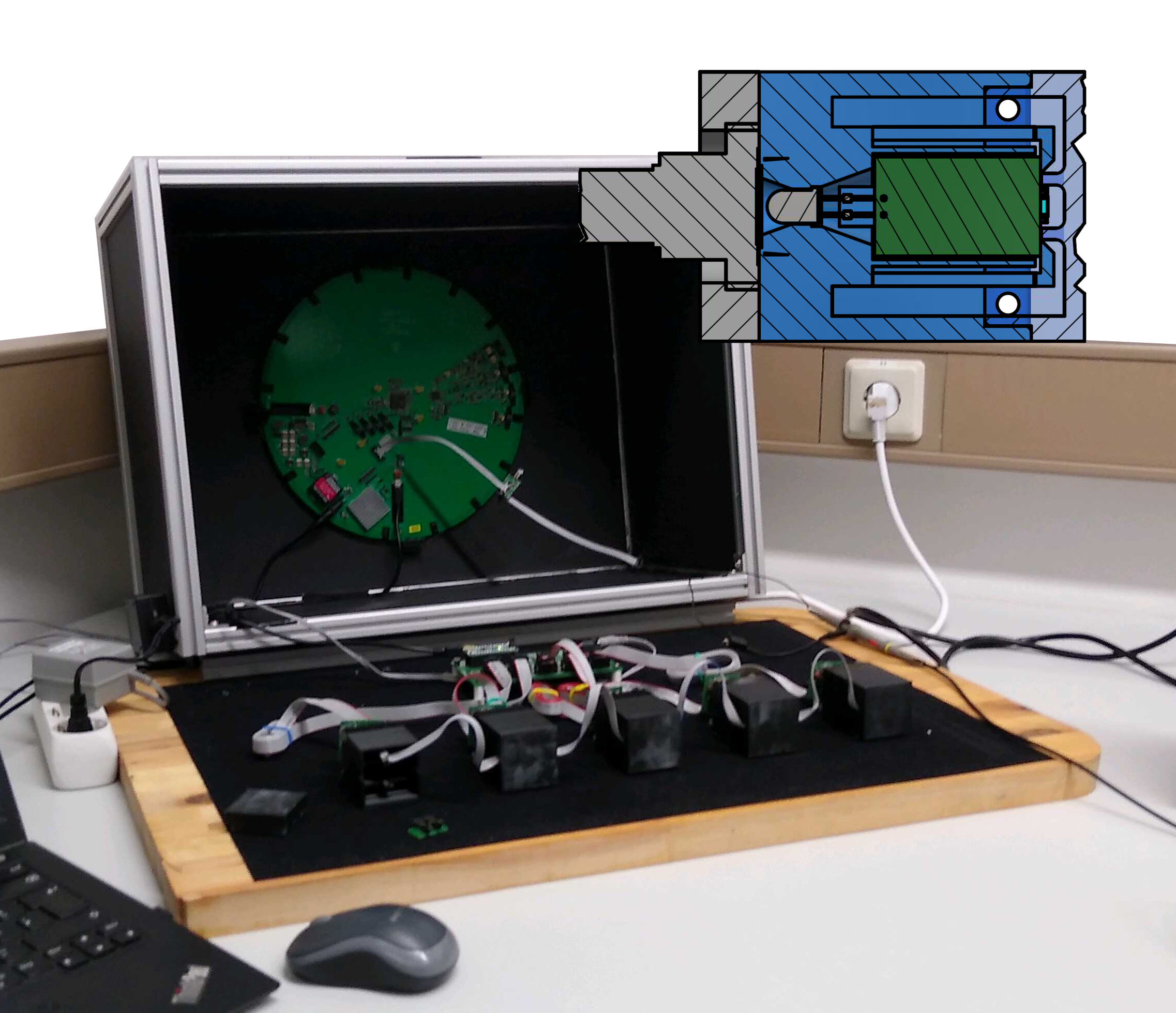}
    }\hfill
    \subfloat[Close-up of the D-Egg testing jig. The annular PCB slots into a 3D-printed structure such that each LED precisely faces its respective readout photodiode.\label{subfig-3:deggtesting}]{%
      \includegraphics[width=0.3\textwidth]{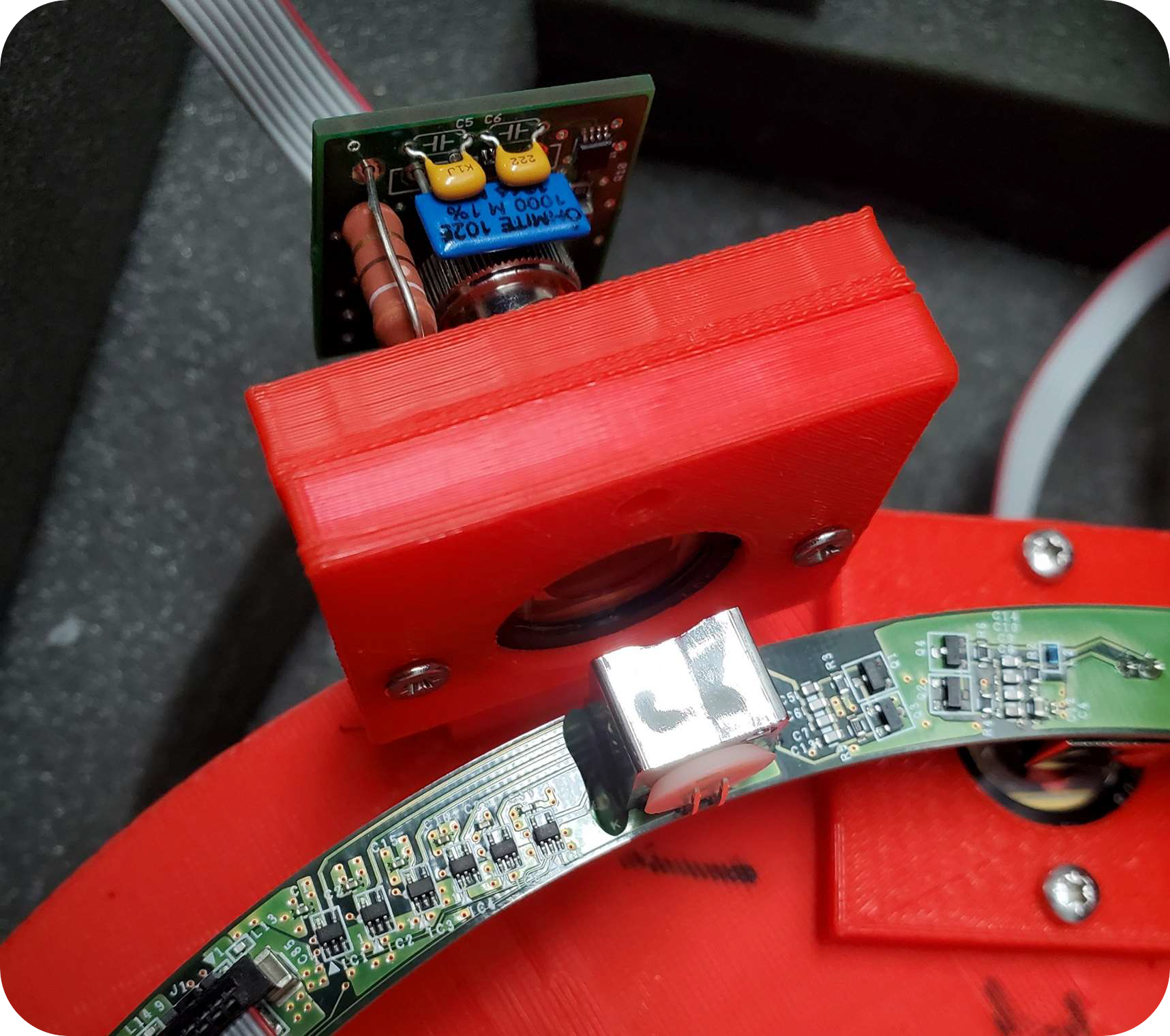}
    }
    \caption{The photodiode systems used for final acceptance testing.}
\end{figure}

Not only are the D-Egg and mDOM flasher systems based on the same design, but their respective testing setups have also been built on common instrumentation to facilitate a good comparability of the measurements. 
Each testing setup consists of one Hamamatsu S2281 photodiode per LED in the respective daisy chain. Their signal is read out by a custom picoammeter, following the design described in \cite{Tosi2015, Rongen2018}. It consists of a preamplifier board directly mounted to each photodiode and a central digitizer board as seen in figure \ref{subfig-1:pdsystem}. The relative detection efficiency of all photodiode and preamplifier pairs was evaluated by measuring the same flasher element. All channels were found to be within 2\% of the average efficiency.

For the mDOM testing system, photodiodes are individually mounted inside light-tight plastic enclosures that also fit an LED element as in the mDOM holding structure. Five of these photodiode modules are placed next to each other in a dark box as seen in figure \ref{subfig-2:mDOMtesting}. 
For the D-Egg testing system, the photodiodes are mounted on a jig onto which the annular flasher PCB is placed for testing. A close-up picture of the D-Egg testing jig can be seen in figure \ref{subfig-3:deggtesting}.\\

Due to the LED's spot size in the near field and the angular emission profile, the full LED intensity is not contained in the active area ($\sim 100$\,mm$^2$) of the photodiode in either testing system. Through long-exposure photography with LEDs illuminating a screen at precisely controlled distances, the fraction of light incident on the photodiode active area has been established to be roughly 60\%. If not stated otherwise, requirements and results are given with respect to the intensity detected by the photodiodes. Due to the different distances of the photodiode to the LED, the D-Egg test system is expected to register $86\pm4\%$ of the intensity received by the mDOM testing system. This value was determined from photographic evaluations of the beam profile as a function of distance. 

\subsection{Testing procedures}

With a daisy chain loaded into the testing setup and its dark box closed, testing proceeds through the following sequence:
\begin{enumerate}
    \item \textbf{Setup self-test:} The setup is tested for light tightness by measuring the dark currents with no bias voltage and no triggers going to the flashers.
    \item \textbf{Flasher functionality:} All LEDs are enabled at maximum brightness, ensuring that all flashers are functional. In the case of a broken LED element, the test will terminate and request that the operator swap the broken element and conduct a retest. Fewer than 2\% of the LED elements were rejected.
    \item \textbf{Flasher addressability:} The LEDs are enabled in sequence, and the test ensures that only the corresponding photodiode sees a signal. This step tests the trigger interfaces and ensures addressability.
    \item \textbf{Flasher brightness:} A threshold scan is performed, measuring the maximum brightness and threshold voltage of each LED.
\end{enumerate}

  \begin{figure}[!ht]
  \centering
    \subfloat[An example measurement of the number of photons per second, as calculated from the photocurrent, as a function of pulse frequency at one bias voltage. The slope yields the number of photons per pulse.\label{subfig-1:linearity}]{%
      \includegraphics[height=0.31\textwidth]{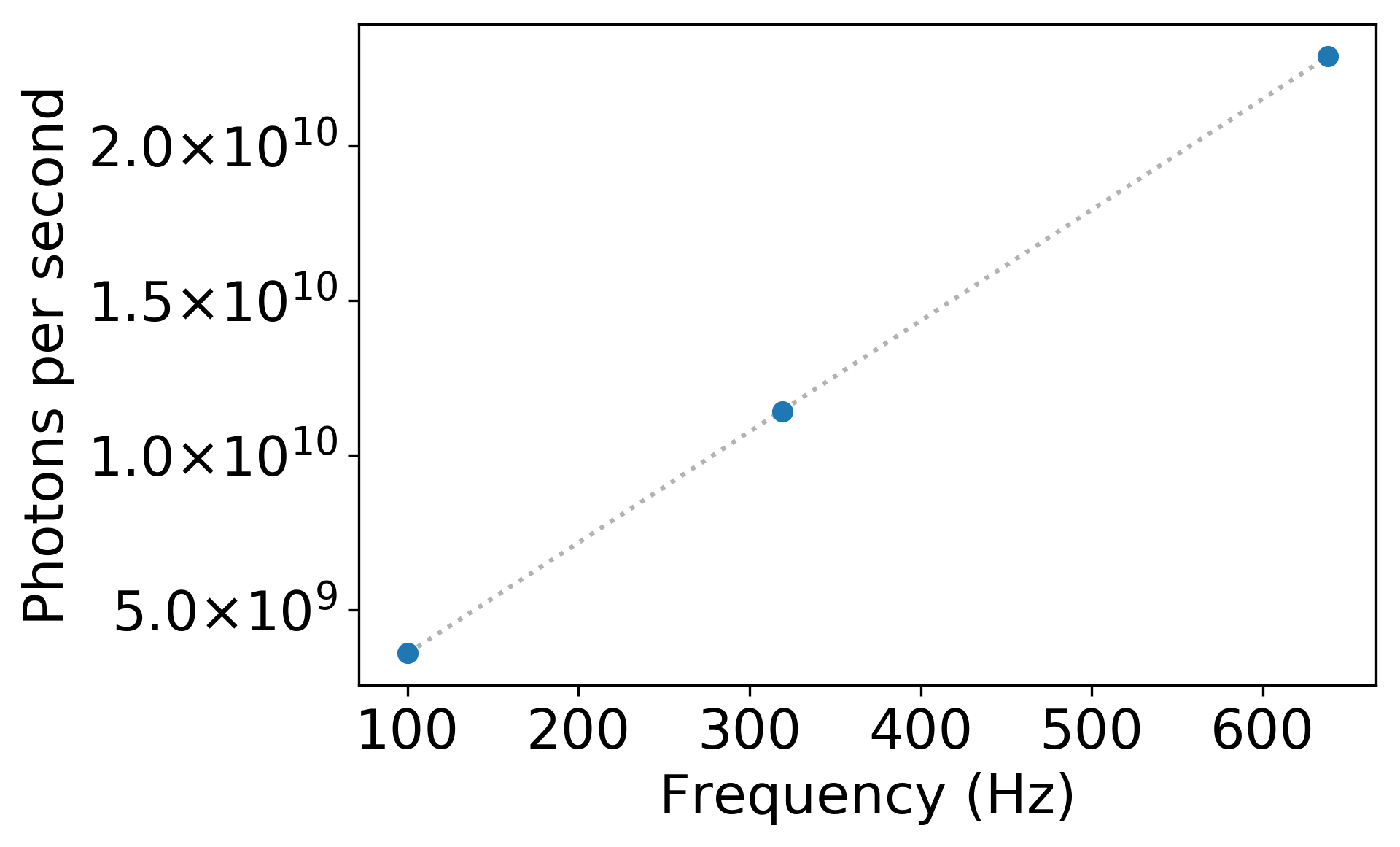}
    }\hfill
    \subfloat[Per-pulse photon yield as a function of bias voltage. The photon count at the maximum bias voltage (red point), as well as the threshold voltage where $5\cdot10^6$ photons per pulse are reached, are recorded.\label{subfig-2:examplescan}]{%
      \includegraphics[height=0.31\textwidth]{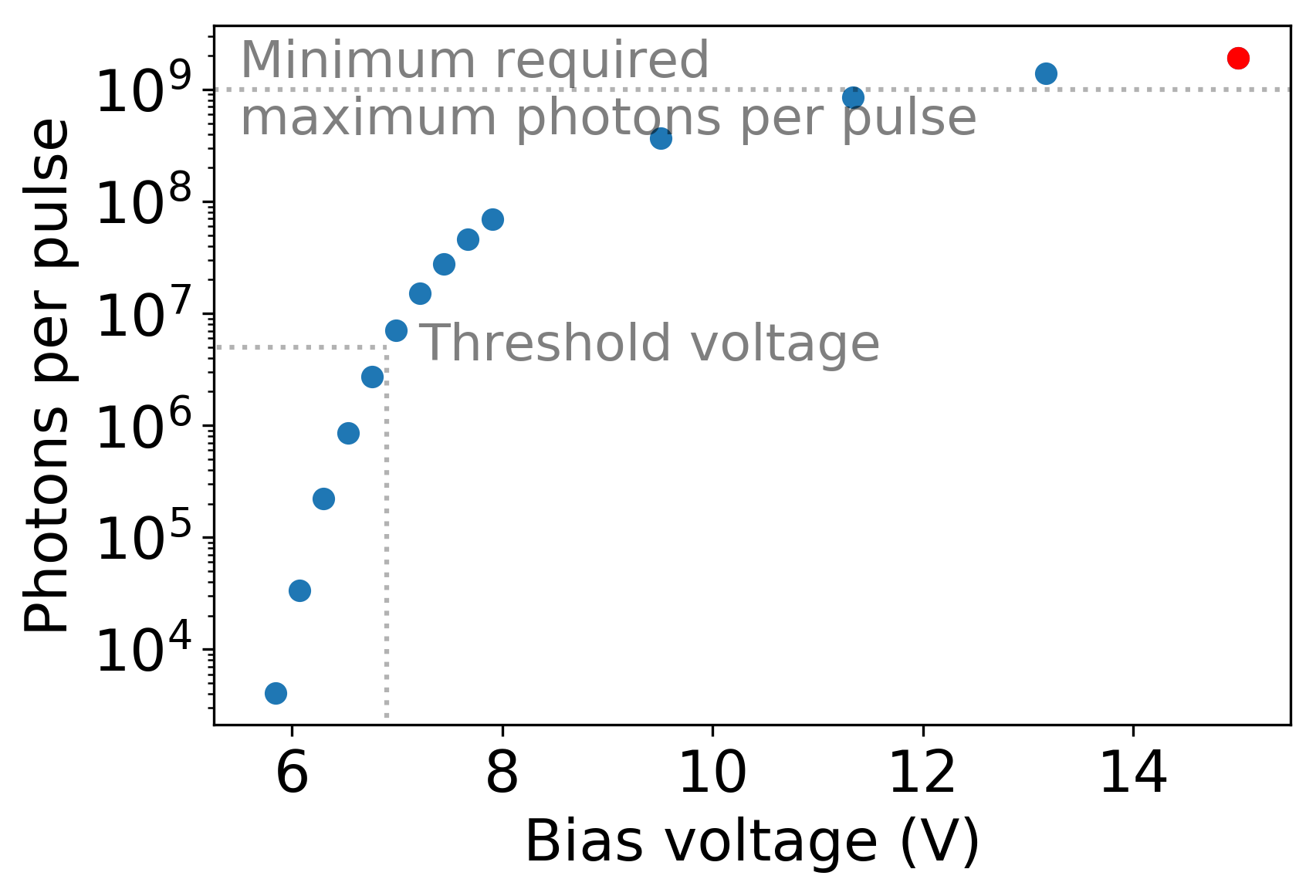}
    }
    \caption{Example measurement of per-pulse photon yield as a function of bias voltage as performed for each LED element.}
  \end{figure}

The threshold scan measures the per-pulse photon yield of each LED at several bias voltages.
At each bias voltage, the photocurrents received at all photodiodes are recorded for at least three repetition frequencies.
Each photocurrent is converted into the equivalent number of measured photons per second taking into account the elementary charge and the datasheet quantum efficiency of the photodiode at 405\,nm. The per-pulse photon count is the slope of a linear fit of the photons per second versus repetition frequency (see figure \ref{subfig-1:linearity}). The repetition frequency starts at 100\,Hz. Should any of the LEDs already saturate at 100\,Hz, the data are discarded and the initial frequency is reduced to 10\,Hz. From the initial photocurrent, two additional frequencies are selected to span the full dynamic range of the photodiode system without risking saturation. 

After recording the photon yield at the maximum bias voltage of 15\,V, the bias voltage is reduced in steps of 1.8\,V to approach the threshold voltage. When the photon count has reduced below the target of $5\cdot10^6$ photons per pulse, the region around the threshold voltage is scanned in steps of $\sim230$\,mV, such that at least three points above the threshold voltage are included and that the lowest recorded per-pulse photon count is below $10^6$. 

An example intensity scan can be seen in figure \ref{subfig-2:examplescan}. This testing procedure on average requires 20 minutes, with the runtime mostly depending on the spread in intensity between the LEDs.

\subsection{Testing results}

  \begin{figure}[!ht]
  \centering
    \subfloat[Histograms of measured maximum photon yields for all D-Egg and mDOM LEDs. The arrows indicate the mean maximum photon count per individual LED position. The D-Egg LEDs feature a position dependence due to layout differences. This is not the case for the mDOM flashers, which are all produced equally, but their distribution is bimodal due to LED characteristics. \label{subfig-1:maxintensities}]{%
      \includegraphics[width=\textwidth]{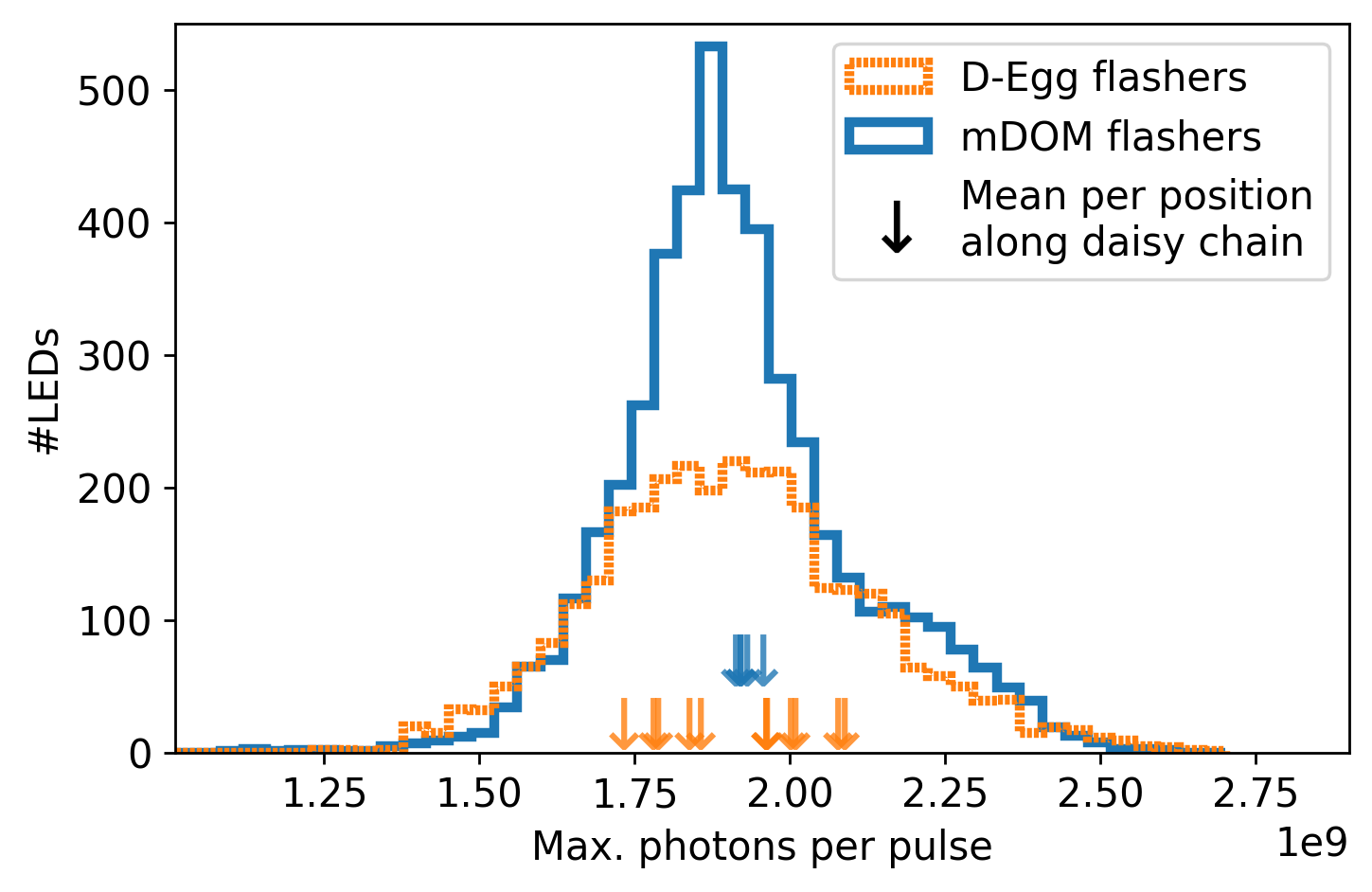}
    }\hfill\vspace{5mm}
    \subfloat[Layout examples for two daisy chain positions on the D-Egg flasher PCB. The top example features longer traces connecting the transistors to the LED, resulting in slightly dimmer emission.\label{subfig-2:deggproblem}]{%
      \includegraphics[width=0.40\textwidth]{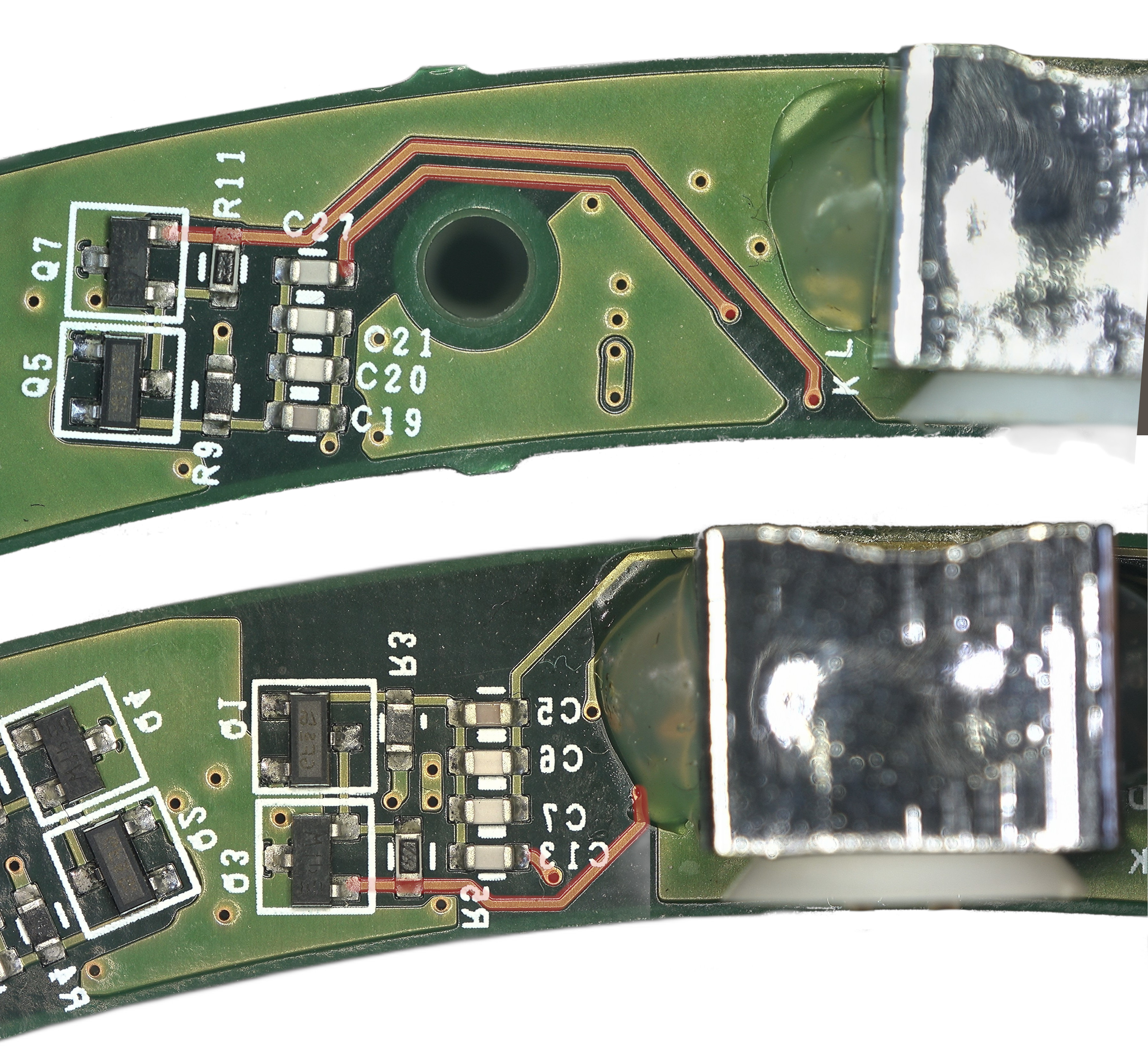}
    }\hfill
    \subfloat[Maximum photon yield as a function of production sequence for the mDOM flashers. LEDs were delivered in bags of 100 each. LEDs from 10 bags feature an average increased maximum brightness.\label{subfig-3:mdomproblem}]{%
      \includegraphics[width=0.55\textwidth]{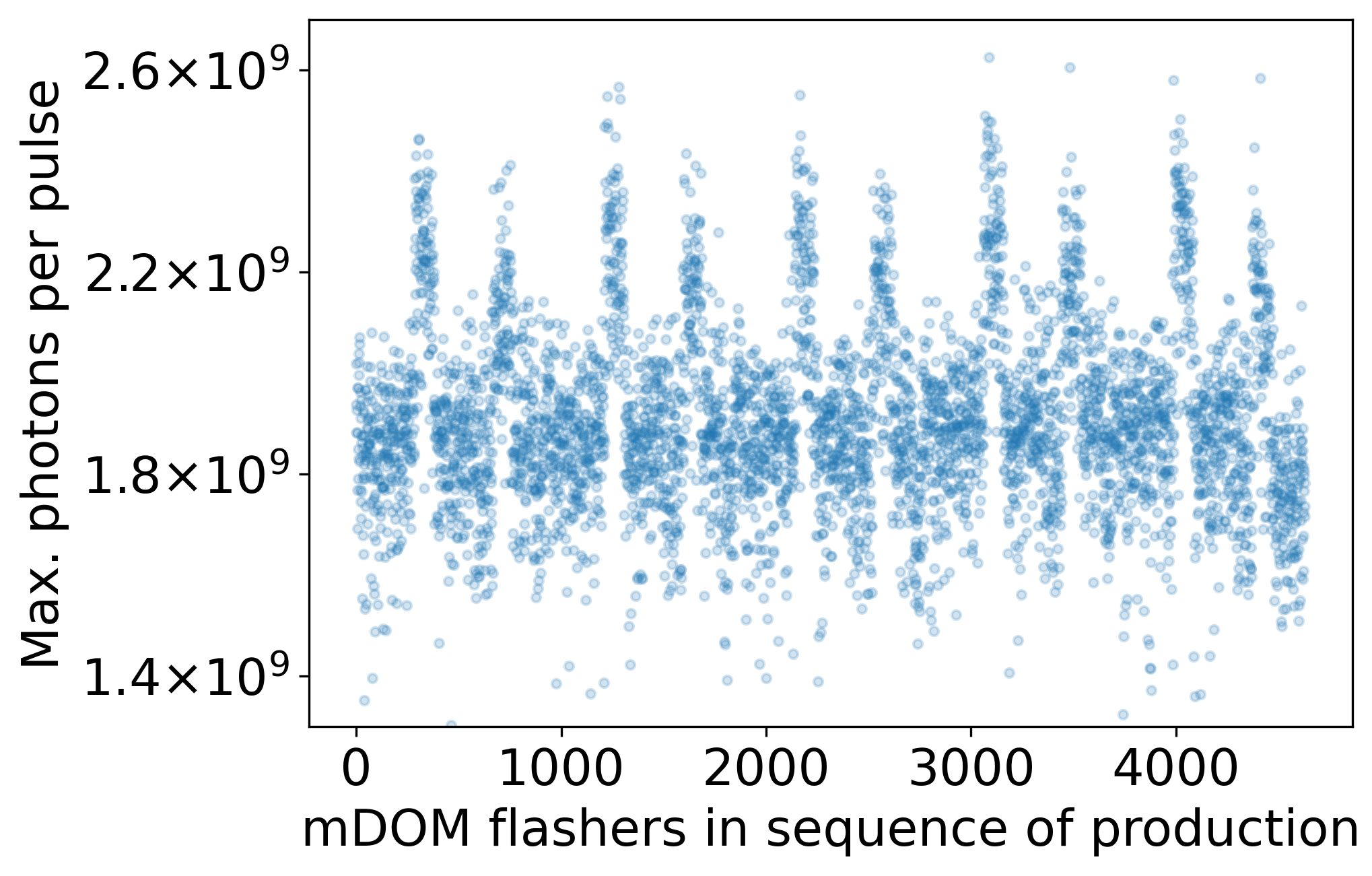}
    }
    \caption{Investigations into the maximum photon yield for both LED calibration systems.\\\,}
  \end{figure}

Figure \ref{subfig-1:maxintensities} shows histograms of the measured maximum photon yields for all D-Egg and mDOM LEDs. The D-Egg values have been corrected by the average relative difference to the mDOM intensities, which is a factor of $1.22=1/0.82$. This correction factor is in good agreement with the expected relative sensitivity of the D-Egg testing system compared to the mDOM testing system as detailed in section \ref{sec:testsetup}. The arrows indicate the mean maximum photon counts for the individual LED positions along the daisy chains. After reworking non-functional LED elements, all tested LEDs conform to the requirement of at least $10^9$ photons at maximum bias voltage. 

The mDOM and D-Egg LEDs show a 10.0\% and 11.6\% standard deviation spread in intensity respectively. For the D-Egg LEDs a significant dependence of the mean photon yield, ranging from $1.7\cdot10^9$ to $2.1\cdot10^9$, on the position along the daisy chain is observed. Correlating the mean intensities to the trace lengths between the driving circuit and the LED shows a clear decrease in intensity with increased trace lengths. Two examples of different trace lengths are seen in figure \ref{subfig-2:deggproblem}. 

Given that the mDOMs use identical PCBs for all LED elements, no such position dependence is seen in the mDOM data. However, the overall distribution still features a non-Gaussian shoulder at high photon counts. Figure \ref{subfig-3:mdomproblem} shows the maximum intensities of mDOM LED elements as a function of the production sequence. The higher intensities do not occur randomly but are grouped in sequences of 100 LEDs. The LEDs have been received from the manufacturer in bags of 100, indicating that the LEDs are not from a single but two production batches.

\begin{figure}[!ht]
  \centering
    \subfloat[Per-pulse photon yield as a function of bias voltage for a random subset of 200 mDOM LEDs. The spread between LEDs is reduced after shifting the curves by the individual threshold voltages. \label{subfig-1:200}]{%
      \includegraphics[width=0.50\textwidth]{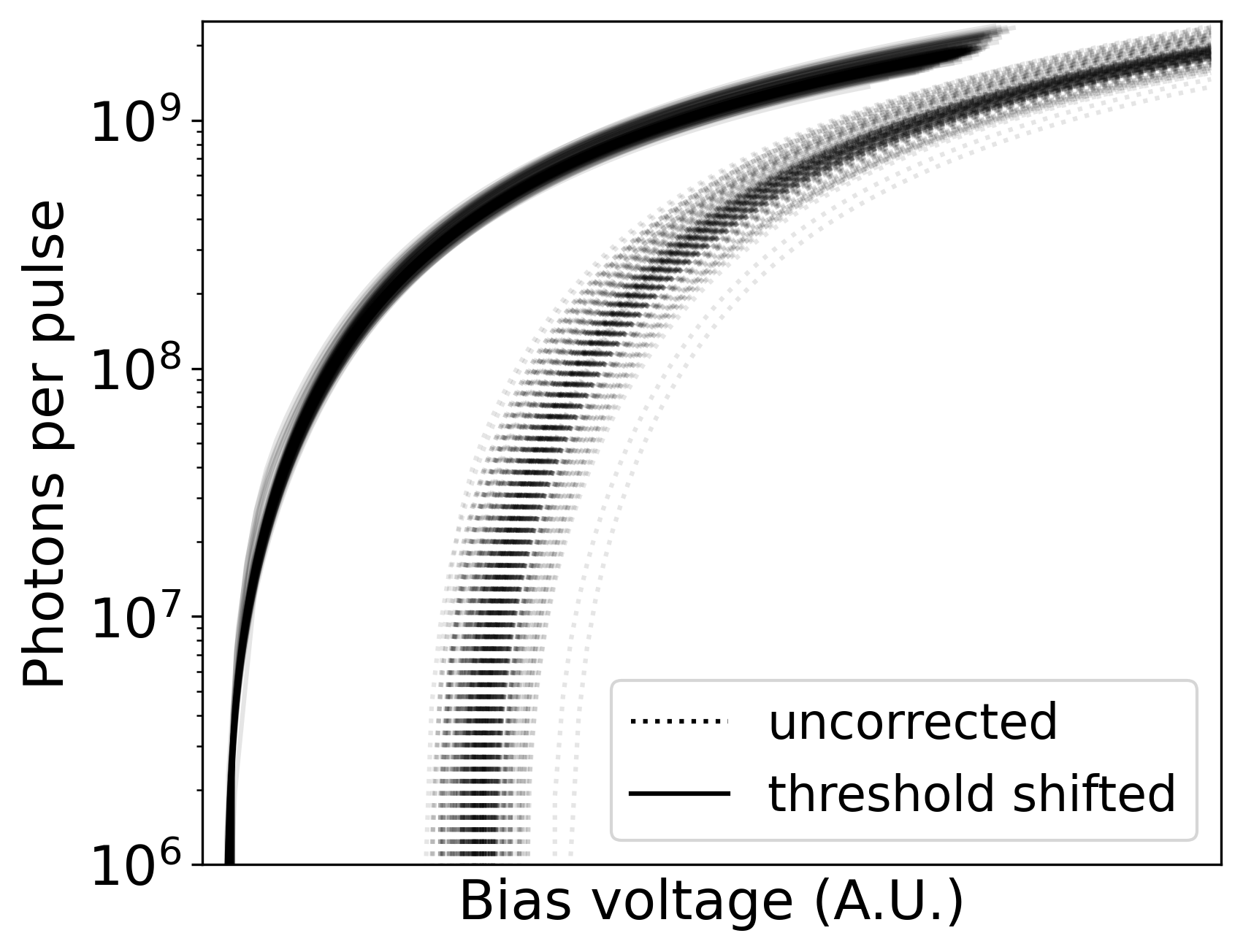}
    }\hfill
    \subfloat[Relative standard deviation (of the photons per pulse over all LEDs divided by the mean intensity) between LEDs as a function of mean intensity given a range of fixed (threshold-corrected) bias voltages. \label{subfig-2:correction}]{%
      \includegraphics[width=0.46\textwidth]{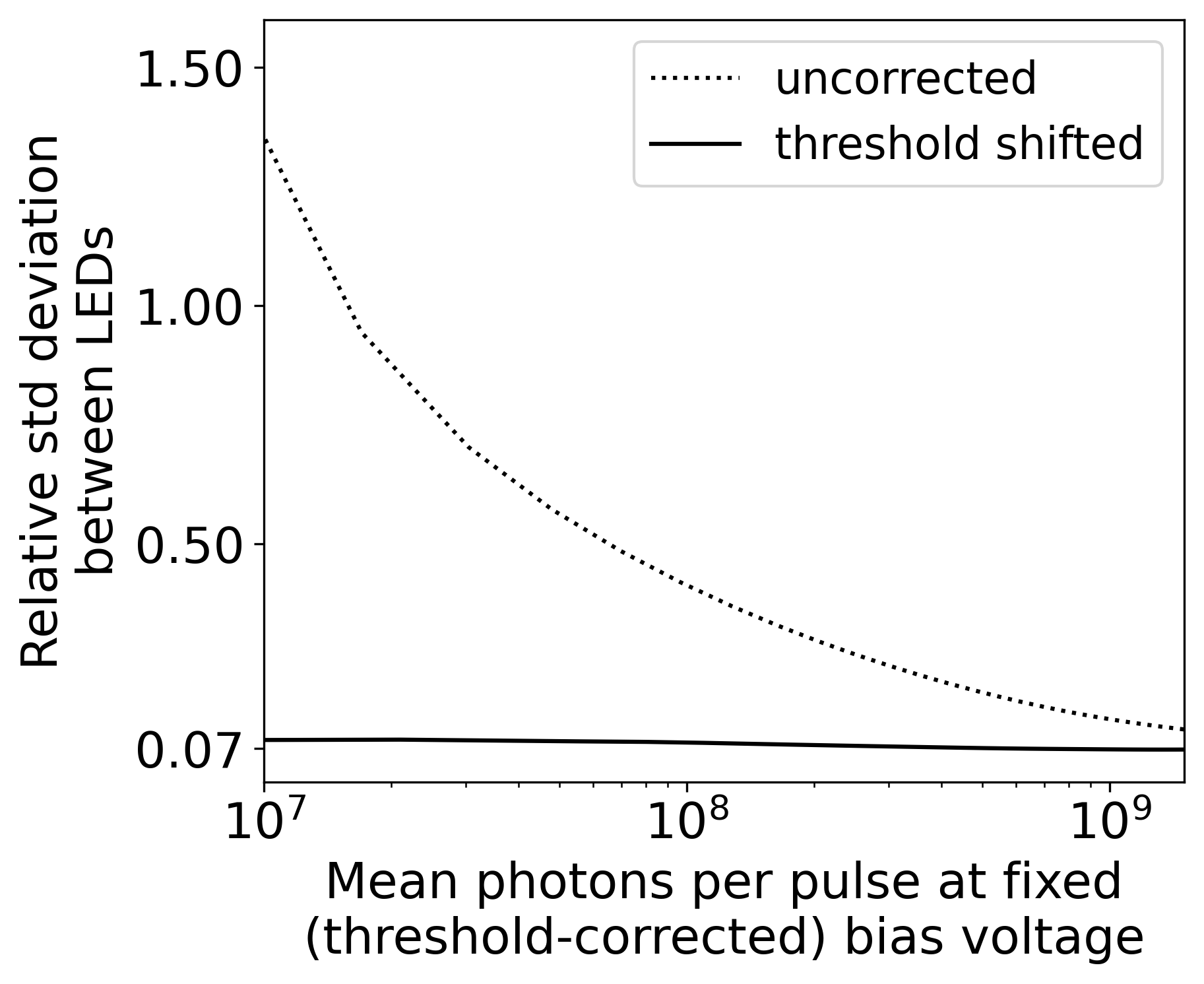}
    }
    \caption{Based on the individually measured threshold voltages, the bias voltage can be configured to yield any desired intensity within 10\% standard deviation.}
  \end{figure}

To illustrate the LED-to-LED fluctuations, the intensity scans of 200 randomly sampled mDOM LEDs are shown in the dotted lines of figure \ref{subfig-1:200}. As discussed above, the spread of intensity is within 10\% between all LEDs of one module type at the highest bias voltage. However, the relative standard deviation (of the photons per pulse over all LEDs divided by the mean intensity at a given bias voltage) quickly grows towards lower intensities. It reaches nearly 150\% at a fixed bias voltage, equating to an average of $10^7$ photons per pulse, as seen in the dotted line in figure \ref{subfig-2:correction}. To facilitate the requirement that the intensity must be configurable to within $\pm 50\%$ of any target value for any LED (see table \ref{tab:requirements}), we must thus operate different LEDs at different bias voltages to obtain the same intensity.
A simple parametrization for the required bias voltages can be obtained by shifting the individual scans by their respective threshold voltages. The solid lines in figure \ref{subfig-1:200} illustrate this approach. This simple correction already reduces the relative standard deviation to 7\% over the full range of bias voltages, and thus well below the requirement, as seen in figure \ref{subfig-2:correction}.\\
At room temperature, these per-LED bias voltages can be interpolated directly from the database of recorded intensity scans. The temperature gradient in the ice will add an additional intensity dependence. Although this remains to be studied in detail, we have already investigated a functional correction where the temperature dependence can be added as a parametrization of the difference between the bias voltage and the threshold voltage.

\section{Use of LEDs in module calibration}

In addition to the 12 LED flashers discussed above, the IceCube Gen1 DOMs feature a dedicated so-called mainboard flasher \cite{detector:paper} that can be dimmed to result in single photon illumination at the PMT of the same module. This setup is used during in situ calibration of the modules. The mDOM and D-Egg designs do not feature such a dedicated flasher. During early testing of the Upgrade flashers in prototype D-Egg and mDOM modules, it was however realized that these can be stably operated below the intended dynamic range, resulting in single photon illumination.

\begin{figure}[!h]
  \centering
    \subfloat[Trigger occupancy as a function of bias voltage. A 10\% occupancy is chosen to select for single photons. \label{subfig-1:FAToccupancy}]{%
      \includegraphics[width=0.48\textwidth]{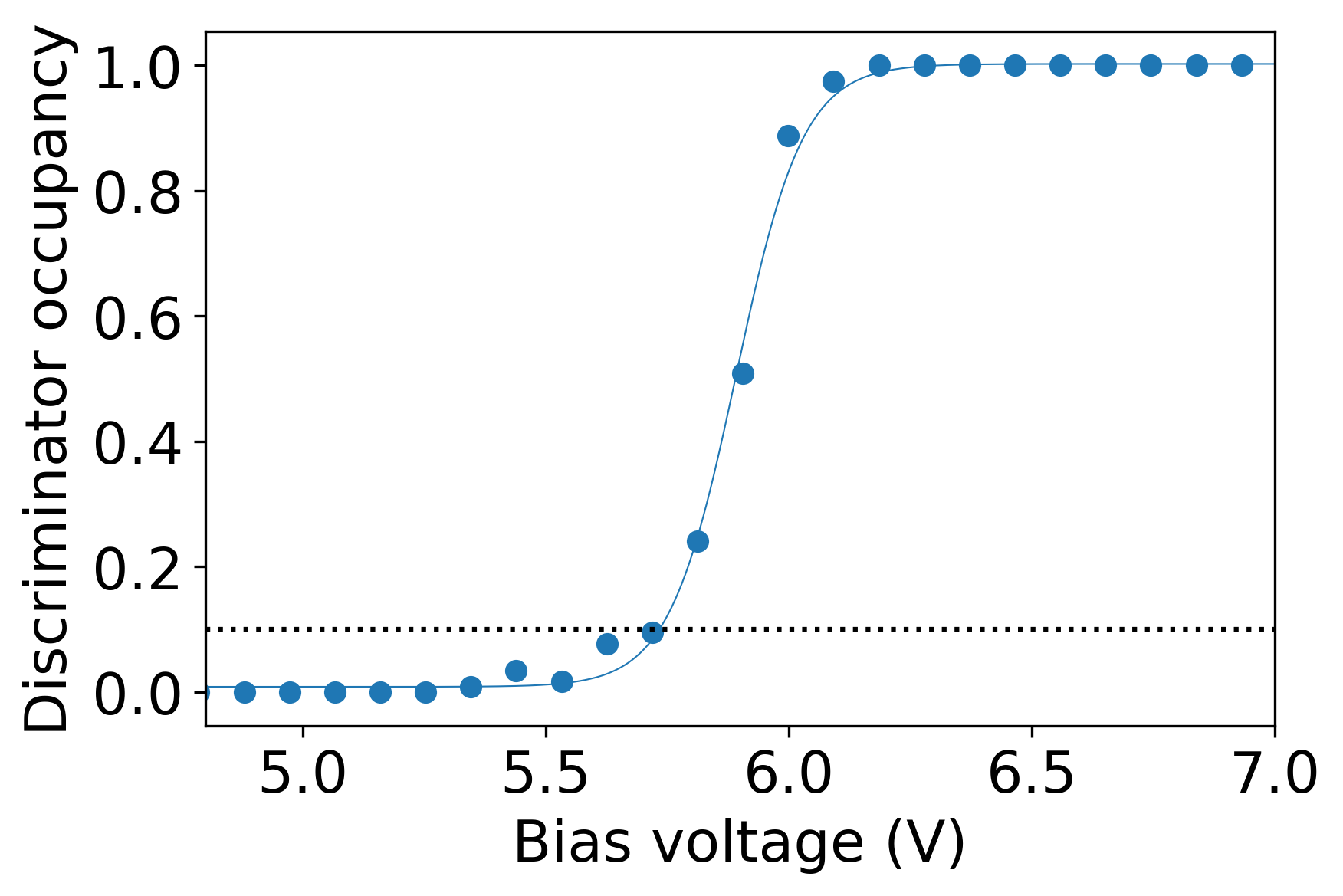}
    }\hfill
    \subfloat[Resulting single photon charge spectrum recorded using the mDOM flasher as a light source.\label{subfig-2:FATPE}]{%
      \includegraphics[width=0.48\textwidth]{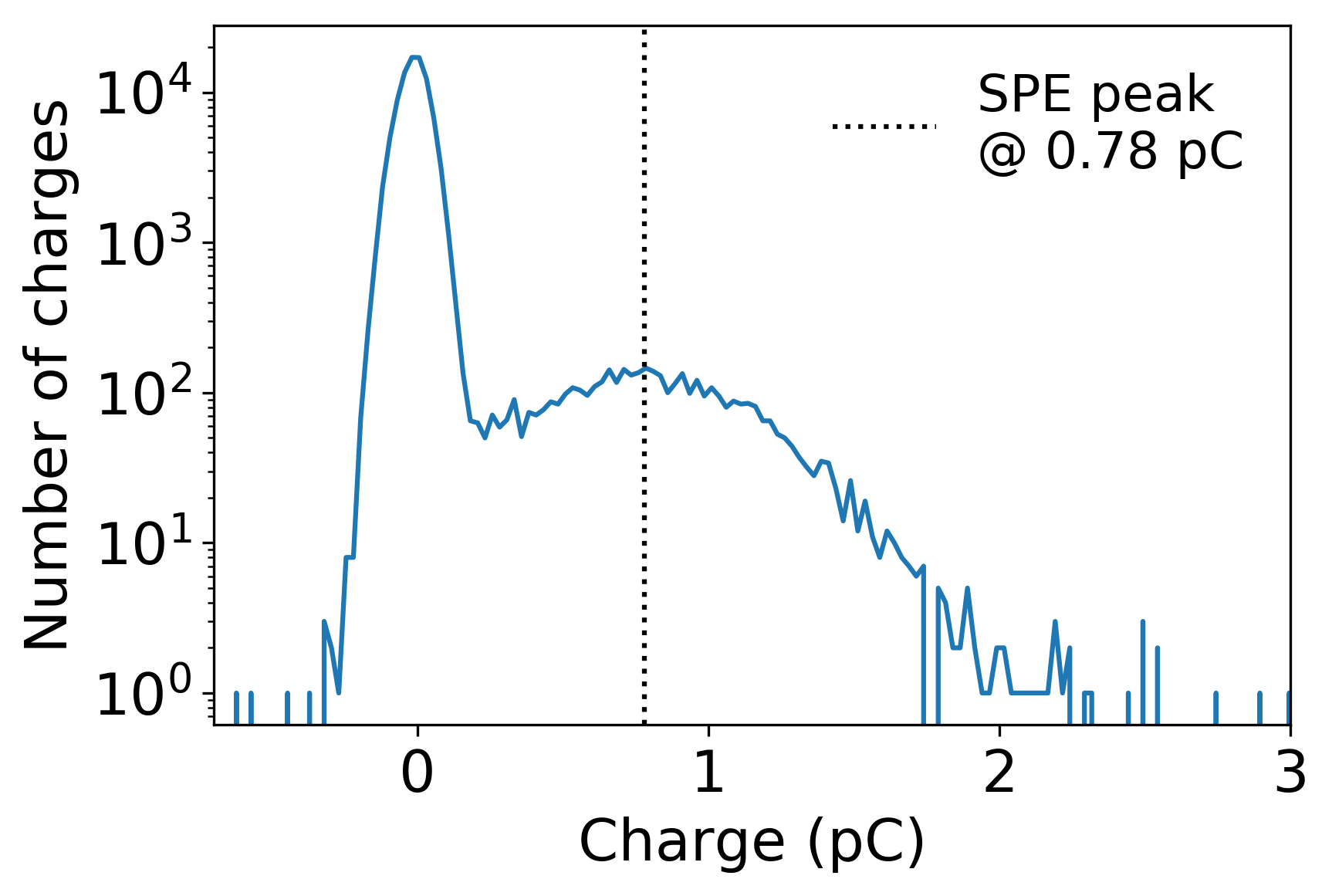}
    }
    \caption{Example operation of an mDOM flasher illuminating a nearby PMT at low bias voltage resulting in a triggered acquisition of low occupancy single photons.}
  \end{figure}

Figure \ref{subfig-1:FAToccupancy} shows the trigger probability for a typical mDOM PMT as a function of the bias voltage of a nearby flasher LED. Figure \ref{subfig-2:FATPE} shows the resulting charge spectrum when the trigger probability is tuned to 10\%, resulting in mostly single photon detection. This configuration allows for easy lab testing (such as final acceptance testing of fully assembled modules) and in situ calibration. 

\section{Conclusion and Outlook}

For Cherenkov neutrino telescopes utilizing natural media, in situ calibration of sensor characteristics and ice/water optical properties often involves the use of pulsed light sources. 
Given the reduced sensor spacing of the Upgrade compared to IceCube, the LED flashers for the IceCube Upgrade sensor modules are dimmer but feature significantly narrower timing profiles. Both systems have been developed in tandem and have been tested to the same specifications with equivalent setups. 
Production and testing of all flasher LEDs as required for the IceCube Upgrade instrumentation have been completed. These are now being integrated into the modules to be deployed during the 2025/26 austral summer season. The performance exceeds the design requirements and is comparable for both module types. Several devices remain in storage and will remain available for long-term testing. This will in particular entail mapping temperature characteristics and their reproducibility between devices. 

\acknowledgments
The authors gratefully acknowledge the support from the following agencies and institutions:
USA {\textendash} U.S. National Science Foundation-Office of Polar Programs,
U.S. National Science Foundation-Physics Division,
U.S. National Science Foundation-EPSCoR,
U.S. National Science Foundation-Office of Advanced Cyberinfrastructure,
Wisconsin Alumni Research Foundation,
Center for High Throughput Computing (CHTC) at the University of Wisconsin{\textendash}Madison,
Open Science Grid (OSG),
Partnership to Advance Throughput Computing (PATh),
Advanced Cyberinfrastructure Coordination Ecosystem: Services {\&} Support (ACCESS),
Frontera and Ranch computing project at the Texas Advanced Computing Center,
U.S. Department of Energy-National Energy Research Scientific Computing Center,
Particle astrophysics research computing center at the University of Maryland,
Institute for Cyber-Enabled Research at Michigan State University,
Astroparticle physics computational facility at Marquette University,
NVIDIA Corporation,
and Google Cloud Platform;
Belgium {\textendash} Funds for Scientific Research (FRS-FNRS and FWO),
FWO Odysseus and Big Science programmes,
and Belgian Federal Science Policy Office (Belspo);
Germany {\textendash} Bundesministerium f{\"u}r Bildung und Forschung (BMBF),
Deutsche Forschungsgemeinschaft (DFG),
Helmholtz Alliance for Astroparticle Physics (HAP),
Initiative and Networking Fund of the Helmholtz Association,
Deutsches Elektronen Synchrotron (DESY),
and High Performance Computing cluster of the RWTH Aachen
and PRISMA Detector Lab at the Johannes Gutenberg University Mainz;
Sweden {\textendash} Swedish Research Council,
Swedish Polar Research Secretariat,
Swedish National Infrastructure for Computing (SNIC),
and Knut and Alice Wallenberg Foundation;
European Union {\textendash} EGI Advanced Computing for research;
Australia {\textendash} Australian Research Council;
Canada {\textendash} Natural Sciences and Engineering Research Council of Canada,
Calcul Qu{\'e}bec, Compute Ontario, Canada Foundation for Innovation, WestGrid, and Digital Research Alliance of Canada;
Denmark {\textendash} Villum Fonden, Carlsberg Foundation, and European Commission;
New Zealand {\textendash} Marsden Fund;
Japan {\textendash} Japan Society for Promotion of Science (JSPS)
and Institute for Global Prominent Research (IGPR) of Chiba University;
Korea {\textendash} National Research Foundation of Korea (NRF);
Switzerland {\textendash} Swiss National Science Foundation (SNSF).


\bibliography{biblio}
\bibliographystyle{JHEP}
\newpage

\end{document}

%% file: authorlist.tex
\author[15]{R. Abbasi,}
\author[62]{M. Ackermann,}
\author[16]{J. Adams,}
\author[38,a]{S. K. Agarwalla,}
\author[9]{J. A. Aguilar,}
\author[20]{M. Ahlers,}
\author[21]{J.M. Alameddine,}
\author[34]{S. Ali,}
\author[42]{N. M. Amin,}
\author[40]{K. Andeen,}
% ...
\affiliation[9]{Universit{\'e} Libre de Bruxelles, Science Faculty CP230, B-1050 Brussels, Belgium}
% ...
\affiliation[15]{Department of Physics, Loyola University Chicago, Chicago, IL 60660, USA}
\affiliation[16]{Dept. of Physics and Astronomy, University of Canterbury, Private Bag 4800, Christchurch, New Zealand}
% ...
\affiliation[20]{Niels Bohr Institute, University of Copenhagen, DK-2100 Copenhagen, Denmark}
\affiliation[21]{Dept. of Physics, TU Dortmund University, D-44221 Dortmund, Germany}
% ...
\affiliation[34]{Dept. of Physics and Astronomy, University of Kansas, Lawrence, KS 66045, USA}
% ...
\affiliation[38]{Dept. of Physics and Wisconsin IceCube Particle Astrophysics Center, University of Wisconsin{\textemdash}Madison, Madison, WI 53706, USA}
% ...
\affiliation[40]{Department of Physics, Marquette University, Milwaukee, WI 53201, USA}
% ...
\affiliation[42]{Bartol Research Institute and Dept. of Physics and Astronomy, University of Delaware, Newark, DE 19716, USA}
% ...
\affiliation[62]{Deutsches Elektronen-Synchrotron DESY, Platanenallee 6, D-15738 Zeuthen, Germany}
\affiliation[a]{also at Institute of Physics, Sachivalaya Marg, Sainik School Post, Bhubaneswar 751005, India}
% ...
